\documentclass[fleqn,usenatbib]{mnras}

\usepackage{ulem}

\usepackage[T1]{fontenc}

\DeclareRobustCommand{\VAN}[3]{#2}
\let\VANthebibliography\thebibliography
\def\thebibliography{\DeclareRobustCommand{\VAN}[3]{##3}\VANthebibliography}

\usepackage{graphicx}	
\usepackage{amsmath}	
\usepackage{amssymb}	
\usepackage{pifont}     
\usepackage{multirow}   
\usepackage{newtxtext,newtxmath}

\newcommand{\cmark}{\ding{51}}%
\newcommand{\xmark}{\ding{55}}%

\usepackage{xcolor,colortbl}
\definecolor{LightGray}{HTML}{DFDFDF}
\definecolor{LightPurple}{HTML}{E8DFFF}
\definecolor{LightGreen}{HTML}{D8EDC5}

\title[Momentum of SNe with cosmic rays]{Momentum deposition of supernovae with cosmic rays}

\author[F. Rodr\'iguez Montero et al.]{Francisco Rodr\'iguez Montero,$^{1}$\thanks{E-mail: francisco.rodriguezmontero@physics.ox.ac.uk}
Sergio Martin-Alvarez,$^{2}$
Debora Sijacki,$^{2}$
\newauthor
Adrianne Slyz,$^{1}$
Julien Devriendt,$^{1,3}$
and Yohan Dubois$^{4}$
\\
$^{1}$Astrophysics, University of Oxford, Keble Road, Oxford OX1 3RH, UK\\
$^{2}$Institute of Astronomy and Kavli Institute for Cosmology, University of Cambridge, Madingley Road, Cambridge CB3 0HA, UK\\
$^{3}$CNRS, Centre de Recherche Astrophysique de Lyon, Universit\'e de Lyon, Universit\'e Lyon 1, ENS de Lyon, UMR 5574, F-69230 Saint-Genis-Laval, France\\
$^{4}$Institut d’Astrophysique de Paris, UMR 7095, CNRS, UPMC University of Paris VI, 98 bis boulevard Arago, 75014 Paris, France\\
}

\date{Accepted XXX. Received YYY; in original form ZZZ}

\pubyear{2021}

\begin{document}
\label{firstpage}
\pagerange{\pageref{firstpage}--\pageref{lastpage}}
\maketitle

\begin{abstract}
The cataclysmic explosions of massive stars as supernovae are one of the key ingredients of galaxy formation. However, their evolution is not well understood in the presence of magnetic fields or cosmic rays (CRs). We study the expansion of individual supernova remnants (SNRs) using our suite of 3D hydrodynamical (HD), magnetohydrodynamical (MHD) and CRMHD simulations generated using {\sc ramses}. We explore multiple ambient densities, magnetic fields and fractions of supernova energy deposited as CRs ($\chi_{\rm CR}$), accounting for cosmic ray anisotropic diffusion and streaming. All our runs have comparable evolutions until the end of the Sedov-Taylor phase. However, our CRMHD simulations experience an additional CR pressure-driven snowplough phase once the CR energy dominates inside the SNR. We present a model for the final momentum deposited by supernovae that captures this new phase: $p_{\rm SNR} = 2.87\times 10^{5} (\chi_{\text{CR}} + 1)^{4.82}\left(\frac{n}{\text{cm}^{-3}}\right)^{-0.196} M_{\odot}$ km s$^{-1}$. Assuming a 10\% fraction of SN energy in CRs leads to a 50\% boost of the final momentum, with our model predicting even higher impacts at lower ambient densities. The anisotropic diffusion of CRs assuming an initially uniform magnetic field leads to extended gas and cosmic ray outflows escaping from the supernova poles. We also study a tangled initial configuration of the magnetic field, resulting instead in a quasi-isotropic diffusion of CRs and earlier momentum deposition. Finally, synthetic synchrotron observations of our simulations using the {\sc polaris} code show that the local magnetic field configuration in the interstellar medium modifies the overall radio emission morphology and polarisation.
\end{abstract}




\section{Introduction}\label{sec:introduction}
Massive stars undergo processes throughout their life cycle that deposit energy and momentum in their vicinity. Their cataclysmic explosions as supernovae (SNe) are central to galaxy formation and evolution, where SNe are often invoked as a central feedback mechanism, particularly dominant at the low-mass end of the galaxy population (see e.g. \citealt{Naab2016,Vogelsberger2020}). Without the energy injection from SNe into the ISM, the runaway gravitational collapse of gas \citep{Kay2002,Tasker2011,Hopkins2011,Dobbs2011} turns the majority of baryons into stars \citep{Somerville1999,Springel2003}, in clear disagreement with observations \citep[e.g.,][]{Behroozi2013}. In order to regulate star-formation, SNe need to be efficient enough to drive galactic winds that eject gas from galaxies. Observations have shown that energetic large-scale outflows from galaxies are ubiquitous (see \cite{Veilleux2005} for a review on galactic outflows), which provides an explanation for the observed metal enrichment of the intergalactic medium (IGM) \citep{Booth2012}. Modelling these stellar and active galactic nuclei (AGN) feedback mechanisms in simulations requires following physics from sub-pc scales all the way to the intergalactic scales above kpc. As this cannot be captured in one single simulation, galactic and cosmological simulations rely on sub-grid models.

However, different simulations have showcased the inefficiency of SN feedback to halt star formation, which has lead to the development of alternative models with the aim to provide closer results to observations (see e.g. \citealt{Rosdahl2018, Smith2018} for a comparison of methods). Because SNe primarily occur within regions of cold and dense gas where radiative cooling is more effective, the low resolution of cosmological simulations is not enough to adequately resolve this phase \citep{Vogelsberger2020}. Solutions to this non-physical over-cooling include disabling the cooling of the affected gas for a limited amount of time \citep[e.g.][]{Stinson2006,Teyssier2013}, employing a stochastic energy deposition \citep[e.g][]{DallaVecchia2012}, or applying gas `kicks' in analytic momentum-driven winds \citep{Murray2005, Kimm2015}. The use of hydrodynamically decoupled wind particles is required in many large volume cosmological simulations to reproduce realistic outflows \citep{Springel2003, Oppenheimer2009, Vogelsberger2013, Pillepich2018}. The attention has now been shifted towards the study of physical components unaccounted for by simulations that have the potential to affect the deposition of energy and momentum by SNe. Feedback agents like radiation \citep{Hopkins2011,Aumer2013,Rosdahl2018,Emerick2018}, Lyman-$\alpha$ pressure \citep{Dijkstra2008,Kimm2018} or cosmic rays (CRs) \citep[e.g][]{Wadepuhl2011,Uhlig2012,Salem2014,Hopkins2021} are explored in addition to the suggested better treatment of star-formation \citep{Kimm2015,Hopkins2018,Semenov2018,Martin-Alvarez2020} and the physics of the ISM \citep[][]{Smith2019}.

Multi-wavelength signatures of CR acceleration have been observed in supernovae remnants (SNR) \citep[e.g][]{Vink2009,Dermer2013,Thoudam2012}, suggesting that approximately 10\% of the explosion energy is converted into CR energy \citep{Morlino2012,Helder2013}. Furthermore, the contribution to the energy density from CRs in the Solar Neighbourhood is on the order of the magnetic, thermal and gravitational energies \citep[e.g][]{Webber1998,Ferriere2001}. This apparent equipartition positions CRs as a non-negligible ingredient in the dynamics of the Galactic disk. In fact, evidence of high CR energy density has been found in regions of high star formation in other galaxies \citep{Acciari2009,Persic2010,Mulcahy2014}, as well as a tight correlation between the radio synchrotron emission of CRs and the far-infrared emission of highly obscured star formation in spiral galaxies \citep{Liu2010}.

As CR particles are effectively collisionless, their interactions with the thermal gas component are dictated by the ambient magnetic field and kinetic instabilities \citep{Zweibel2017}. CRs excite Alfvén waves via the streaming instability while travelling along magnetic field lines, from which they scatter reducing their bulk velocity. Through this interaction, CRs can effectively transfer energy and momentum to the thermal gas \citep{Kulsrud1969}. This is the underlying assumption in the macroscopic description of CRs as a relativistic fluid with a softer adiabatic index ($\gamma_{\text{CR}}=4/3$, instead of the $\gamma=5/3$ for a monatomic thermal gas). The direct consequences are that: i) CRs experience lower adiabatic losses due to expansion, ii) the cooling timescales are longer, so they maintain their energy density for a longer period of time, and iii) their rapid diffusion along magnetic field lines enables CRs to escape dense and compact regions, reaching the diffuse gas where they have the potential to drive fast and steady galactic outflows \citep[e.g.][]{Everett2008}.

Based on the acceleration of CRs by SNe shocks, the implementation of CRs in galaxy and ISM simulations has been generally done by employing SNe as sources of CR energy density \citep[see also e.g.,][for implementations in AGN feedback]{Sijacki2008, Ruszkowski2017, Ehlert2018}. These models inject a fraction of the original SNe energy in the form of CR energy to the surrounding gas in addition to the original thermal and/or kinetic feedback mechanism of each SN. After injection, CRs are evolved following CR dynamics. In early works, streaming at the local sound speed and isotropic diffusion was the way in which CR transport was implemented \citep[e.g.][]{Jubelgas2008,Uhlig2012}. This showed that including CR as a separate fluid provided a mechanism to drive powerful winds in low mass galaxies \cite{Uhlig2012} with high mass loading factors \citep{Salem2014}, and thickened gaseous disks \citep{Booth2013}. However, a more realistic treatment of CR transport has recently become possible largely due to advances in MHD codes. Multiple new implementations have been developed for the physics describing CR anisotropic diffusion and streaming, their coupling with the thermal gas and cooling due to hadronic, Coulomb and ion losses \citep[e.g.][]{Girichidis2014,Dubois2016,Chan2019}, as well as their acceleration at shocks \citep{Pfrommer2017,Dubois2019}. In state-of-the-art CR numerical methods, it is now common to find that including CRs results in the formation of powerful and steady outflows \citep{Ruszkowski2017,Girichidis2018,Hopkins2021}, the suppression of star formation \citep{Jubelgas2008,Pfrommer2017,Hopkins2021,Dashyan2020}, or the ejection of large fractions of the baryons found in the galaxies out to their virial radii \citep{Pfrommer2007,Pakmor2016}. Overall, simulations that included CR transport and studied their effect on wind launching found that CRs increased the mass loading of winds, making them also colder and denser.

Despite their obvious importance, the effect of CRs on the evolution of SNRs is not yet fully understood. Early implementations of CR pressure in the solution of the blast wave problem \citep{Chevalier1983} suggested that its addition does not cause the evolution of supernova remnants to significantly deviate from the self-similar solution. More recently, numerical methods that include CR physics have been used in three dimensional simulations of spherical blast waves \citep{Pfrommer2017,Pais2018a,Dubois2019}, suggesting instead a Sedov-Taylor solution with an effective adiabatic index of $7/5$. \citet{Girichidis2014} studied the dynamical impact of CRs during the early adiabatic phase of the expansion of individual SNRs using their implementation of anisotropic transport of CRs. They found an effective acceleration of cold gas in the ambient medium above that expected from the thermal solution. Furthermore, since CRs have longer cooling times, they undergo significantly lower radiative losses than thermal gas during the pressure-driven snowplough phase. This provides an extra source of energy that fuels further expansion \citep{Diesing2018}. Hence, the inclusion of CRs in SNR modelling could affect the final energy and momentum deposition. Additionally, the nature of magnetic field structure in SNRs has been highly debated \citep[see][for a review]{Reich2002}, since observations of synchrotron polarised emission suggest a transition from a mainly turbulent field in early SNRs \citep{Gaensler1997,DeLaney2002}, to a tangential field in older ones \citep{Reynolds1993}. Modelling the effects of CRs in different magnetic fields within SNRs could help understanding these polarisation signatures and account for the low ($\sim 2-8$\%) polarisation fraction \citep[e.g.][]{Reynoso2013,Reynoso2018}.

To address these issues, in this work we perform a detailed study to quantify the effects of magnetic fields and CRs on the dynamical evolution of SNRs expanding in a homogeneous medium with a representative range of densities. We explore different magnetic field strengths as well as configurations, including uniform and tangled initial conditions. We account for anisotropic diffusion and streaming of CRs where we vary the initial energy fraction residing in CRs as well. This paper is organised as follows: in Section~\ref{sec:methods_ics} we explain the CRMHD numerical methods employed and our choice of initial conditions that explore a significant region of the parameter space where SNe are typically found. In Section~\ref{sec:results_align} we study the separate effects of magnetic fields and CRs for an aligned magnetic field, proposing an updated model for the momentum deposition of SNe that captures the effects of CRs. In Section~\ref{sec:results_tangled} we explore an initially tangled magnetic field configuration instead. Section~\ref{sec:mocks} reviews the observational signatures of CRs and the magnetic field configuration in mock synchrotron observations. Finally, in Sections~\ref{sec:discussion} and \ref{sec:conclusions} we present the discussion and main conclusions of our work.

\section{Numerical methods and Initial conditions}\label{sec:methods_ics}

In this work we run simulations of individual SNe generated using the MHD code {\sc Ramses} \citep{Teyssier2002,Fromang2006} \footnote{\url{https://bitbucket.org/rteyssie/ramses/src/master/}}, extended by \cite{Dubois2016,Dubois2019} to model CRs. We make use of the Lax-Friedrichs method in order to solve the equations of hydrodynamics (HD), magneto-hydrodynamics (MHD) and cosmic-rays-magneto-hydrodynamics (CRMHD), and a MinMod slope limiter to reconstruct the interpolated variables. 

As we are particularly interested in the late time evolution of the SNR as well as its deposited momentum in the ambient medium, we focus on resolving the end of the Sedov-Taylor, the pressure snowplough and the momentum snowplough phases. We run cubic boxes with $225$~pc length per side for a minimum duration of at least $1$~Myr and up to $6$~Myr. The low level resolution of the grid is fixed to a uniform distribution of $64^3$~cells, which yields a resolution for the ambient medium of $\Delta x \simeq 3.52$~pc. For the AMR we refine on relative gradients higher than 0.2 for gas density, velocity, thermal pressure and CR pressure, reaching a minimum cell size of $0.44$~pc. In addition, we enforce the two neighbouring cells to each refined cell to also be refined. This provides a smoother transition between different levels of resolution around large gradients.
\begin{table}
 \caption{Initial conditions of the uniform ambient medium gas for individual SN simulations, reproducing \protect\cite{Iffrig2014}. From left to right: number density $n$, temperature $T$, and the corresponding plasma $\beta$ for the different magnetic field initial strengths.}
 \label{tab:init_cond_ambient}
 \centering
 \begin{tabular}{ccc}
  \hline
  n [cm$^{-3}$] & T [K] & $\beta$\\
  \hline
  \multirow{3}{*}{1} & \multirow{3}{*}{4907.8} &$135$\\
  &&$1.35$\\
  &&$0.054$\\
  \hline
  \multirow{3}{*}{10} & \multirow{3}{*}{118.16}&$1350$\\
  &&$13.5$\\
  &&$0.542$\\
  \hline
  \multirow{3}{*}{100} & \multirow{3}{*}{36.821} & $13500$\\
  &&135\\
  &&5.42\\
  \hline
 \end{tabular}
\end{table}

Our initial conditions (shown in Table~\ref{tab:init_cond_ambient}) follow those of \cite{Iffrig2014} for uniform media with typical thermodynamical characteristics of the ISM. The radiative cooling of the gas is modelled using interpolated {\sc cloudy} cooling tables \citep{Ferland1998} above $10^4$K and following \cite{Rosen1995} below $10^4$K. The energy of the SN is fixed across all simulations to the canonical value of $E_{\text{SN}}=10^{51}$~erg, distributed amongst the 8 central cells of the box at the start of the simulation. 

When included, magnetic fields are exclusively seeded as part of the initial conditions of our simulations and they are subsequently evolved self-consistently employing ideal MHD. We explore two different types of initial configurations of the magnetic field. The first type is a uniform magnetic field with strength $B_0$ and its direction aligned with the $z$ axis of the computational domain. The second type is a tangled magnetic field that follows a magnetic energy spectrum with spectral index $n_B$. The tangled magnetic initial conditions (ICs; Martin-Alvarez et al. in prep.; but also similar to those employed in \citealt{Katz2021}) are produced by modulating the spectrum of the vector potential at each Fourier wavelength $k$ and requiring the magnetic energy spectra produced by its curl to have a spectral slope $n_B$. All our magnetic ICs assume $n_B = 3/2$ to reproduce a Kazantsev spectrum \citep{Kazantsev1968} corresponding to the magnetic inverse-cascade characteristic of a tangled ISM \citep{Bhat2013,Martin-Alvarez2018}. The computation of our magnetic ICs in the form of the curl of a vector potential field ensures a null divergence by construction. Finally, we renormalise the strength of the magnetic field in these ICs so that the total magnetic energy in the simulated box matches that of our uniform magnetic field runs. We identify the normalisation of each of these sets of ICs, $B_{\rm eff}$, using the value of $B_0$ of the corresponding aligned magnetic field run with the same total magnetic energy. The 8 cells initially containing the SN employ the same type of magnetic field configuration as the rest of the domain (i.e. no magnetic energy is injected with the SN). The observed magnetic field in the ISM and in giant molecular clouds (GMCs) typically ranges from 1 to $\sim 10\,\mu$G for $n=1-100$~cm$^{-3}$ \citep{Beck2001,Crutcher2010}. Consequently, the majority of our suite of simulations explore magnetic fields with strengths of 1 and $5$~$\mu$G (see values of plasma beta in Table~\ref{tab:init_cond_ambient}). We further consider the case of weak magnetisation ($B = 0.1\,\mu$G).

In all our simulations, we assume the bulk of CRs to be accelerated at sub-pc scales, and thus inject a CR energy $E_{\text{CR, inj}} = \chi_{\text{CR}}E_{\text{SN}}$ in the central 8 cells. Here, $\chi_{\text{CR}}$ is the CR fractional contribution to the total energy of the SN. Non-relativistic shocks can channel as much as $10$-$20$\% of their bulk energy into CRs \citep{Caprioli2014}. We explore $\chi_{\text{CR}} = 0.01$, $0.05$ and $0.1$, hence injecting initially total CR energies equivalent to $10^{49}$, $5\times 10^{49}$ and $10^{50}$~erg, respectively. We fix the diffusion coefficient $D_{0}$ to $3\times 10^{27}$~cm$^2$~s$^{-1}$, opting for a low value to account for possible confinement at smaller scales \citep[e.g.][]{Nava2016,Brahimi2019}.

We set the metallicity equal to the solar metallicity, and do not inject additional metals with the SN ejecta. Boundaries have been set as outflowing, i.e. gas that crosses the boundaries escapes the simulation domain. Considering our box size $L_{\rm box}$ of 225 pc, we estimate the earliest escape of CRs at $t_{\rm esc} \simeq(L_{\rm box}/2)^2/D_0 \simeq 1.27$~Myr. The list of all our studied simulations is shown in Table~\ref{tab:sims}.

\begin{table*}\label{tab:sim_list}
 \caption{Summary of all the simulations studied in this work (40 in total). All simulations use a box size of $225$~pc, except the M01HD1CR10b run, marked with an asterisk (\textasteriskcentered), which has a box size of $900$~pc. All simulations have a maximum resolution of $0.44$~pc. Columns show for each simulation: initial ambient particle number density ($n$), initial magnetic field strength ($B_0$), initial fraction of $E_{\text{SN}}$ injected as CR energy ($\chi_{\text{CR}}$), and whether the simulations have been evolved until the total deposited momentum converged during the momentum conserving phase($p_{\rm SNR}$ convergence). Simulations with a dagger symbol (\textdagger) employ tangled magnetic field initial conditions. Our fiducial simulations are highlighted, with gray, purple and green colours indicating HD, MHD and CRMHD, respectively.}
 \label{tab:sims}
 \centering
 \begin{tabular}{lcccc}
  \hline
  Simulation & n & $B_0$ [$\mu$G] & $\chi_{\text{CR}}$ & $p_{\rm SNR}$ convergence\\
  \hline
  HD1 &  1 & 0.0 & 0.0 & \cmark\\
  \rowcolor{LightGray} HD10 &  10 & 0.0 & 0.0 & \cmark\\
  HD100 &  100 & 0.0 & 0.0 & \cmark\\
  \hline
  M01HD1 &  1 & 0.1 & 0.0 & \cmark\\
  \rowcolor{LightPurple} M01HD10 &  10 & 0.1 & 0.0 & \cmark\\
  M01HD100 &  100 & 0.1 & 0.0 & \cmark\\
  M1HD1 &  1 & 1.0 & 0.0 & \cmark\\
  \rowcolor{LightPurple} M1HD10 &  10 & 1.0 & 0.0 & \cmark\\
  \rowcolor{LightPurple} Mag1HD10\textdagger &  10 & 1.0 & 0.0 & \cmark\\
  M1HD100 &  100 & 1.0 & 0.0 & \cmark\\
  M5HD1 &  1 & 5.0 & 0.0 & \cmark\\
  M5HD10 &  10 & 5.0 & 0.0 & \cmark\\
  M5HD100 &  100 & 5.0 & 0.0 & \cmark\\
  \hline
  M01HD1CR1 &  1 & 0.1 & 0.01 & \xmark\\
  M01HD1CR5 &  1 & 0.1 & 0.05 & \xmark\\
  M01HD1CR10 &  1 & 0.1 & 0.1 & \xmark\\
  M01HD1CR10b\textasteriskcentered & 1 & 0.1 & 0.1 & \xmark\\
  \rowcolor{LightGreen} M01HD10CR1 &  10 & 0.1 & 0.01 & \cmark\\
  \rowcolor{LightGreen} M01HD10CR5 &  10 & 0.1 & 0.05 & \xmark\\
  \rowcolor{LightGreen} M01HD10CR10 &  10 & 0.1 & 0.1 & \xmark\\
  \hline
 \end{tabular}
 \begin{tabular}{lcccc}
  \hline
  Simulation & n & $B_0$ [$\mu$G] & $\chi_{\text{CR}}$ & $p_{\rm SNR}$ convergence\\
  \hline
  M01HD100CR1 &  100 & 0.1 & 0.01 & \cmark\\
  M01HD100CR5 &  100 & 0.1 & 0.05 & \cmark\\
  M01HD100CR10 &  100 & 0.1 & 0.1 & \cmark\\
  \rowcolor{LightGreen} M1HD10CR1 &  10 & 1.0 & 0.01 & \cmark\\
  \rowcolor{LightGreen} M1HD10CR5 &  10 & 1.0 & 0.05 & \xmark\\
  \rowcolor{LightGreen} Mag1HD10CR5\textdagger &  10 & 1.0 & 0.05 & \xmark\\
  \rowcolor{LightGreen} M1HD10CR10 &  10 & 1.0 & 0.1 & \xmark\\
  \rowcolor{LightGreen} Mag1HD10CR10\textdagger &  10 & 1.0 & 0.1 & \xmark\\
  M1HD100CR1 &  100 & 1.0 & 0.01 & \cmark\\
  M1HD100CR5 &  100 & 1.0 & 0.05 & \cmark\\
  M1HD100CR10 &  100 & 1.0 & 0.1 & \cmark\\
  M5HD1CR1 &  1 & 5.0 & 0.01 & \xmark\\
  M5HD1CR5 &  1 & 5.0 & 0.05 & \xmark\\
  M5HD1CR10 &  1 & 5.0 & 0.1 & \xmark\\
  M5HD10CR1 &  10 & 5.0 & 0.01 & \cmark\\
  M5HD10CR5 &  10 & 5.0 & 0.05 & \cmark\\
  M5HD10CR10 &  10 & 5.0 & 0.1 & \cmark\\
  M5HD100CR1 &  100 & 5.0 & 0.01 & \cmark\\
  M5HD100CR5 &  100 & 5.0 & 0.05 & \cmark\\
  M5HD100CR10 &  100 & 5.0 & 0.1 & \cmark\\
  \hline
 \end{tabular}
\end{table*}

\section{Results with Uniform Magnetic Fields}\label{sec:results_align}

\begin{figure*}
  \centering \includegraphics[width=\textwidth]{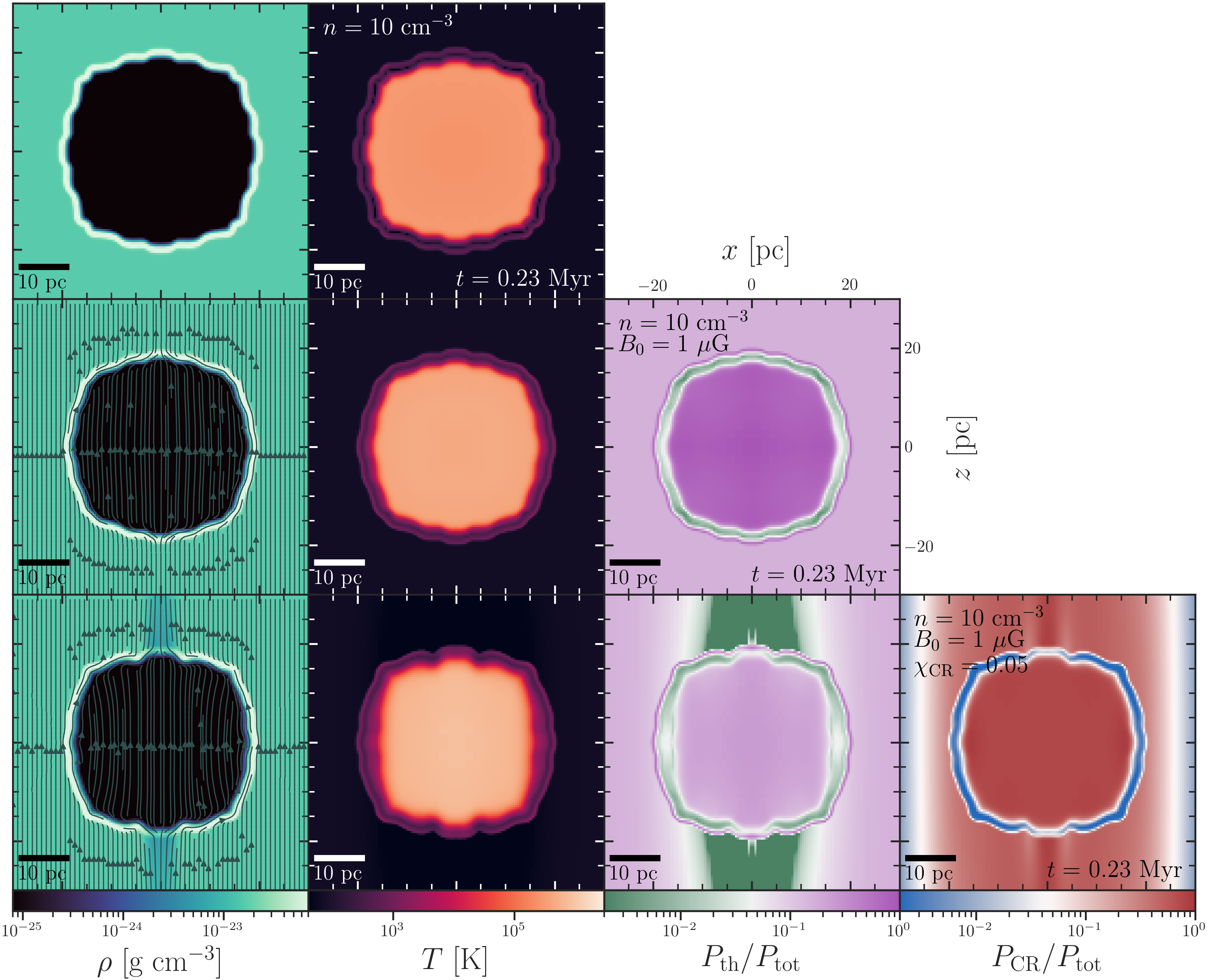}
  \caption{$XZ$ plane slices of the HD (first row), MHD (second row) and CRMHD (third row) SNe simulations, with initial ambient number density $n=10$~cm$^{-3}$, uniform magnetic field $B_0 = 1 \mu$G and the CR energy injection fraction $\chi_{\text{CR}} = 0.05$. \textit{First column}: gas density with overplotted streamlines depicting the orientation of the magnetic field; \textit{second column}: gas temperature; \textit{third column}: ratio of thermal ($P_{\text{th}}$) to total ($P_{\text{tot}}$) gas pressure; and \textit{fourth column}: ratio of CR ($P_{\rm CR}$) to total gas pressure. All slices are shown at $0.25$~Myr. The HD simulation presents the archetypal SNR structure of the shocked, dense shell with an internal low density, high temperature bubble. The SNR is shown well after the end of the Sedov-Taylor phase, when the difference between the ambient and bubble pressure has decreased sufficiently (allowing for the grid-based instabilities of the shell to grow). In the MHD simulation, these instabilities are suppressed due to magnetic pressure. Instead, the figure depicts the onset of shell widening along the $x$ direction (i.e. perpendicular to the initial magnetic field) due to a high magnetic pressure. For the CRMHD simulation, the anisotropic configuration of the magnetic field sets the axial direction as that preferred by escaping CRs, forming elongated cones of high CR energy at the poles. The widening of the shell in the $x$ direction is enhanced due to the additional contribution of CR pressure.}
  \label{fig:uniform_slices}
\end{figure*}

\subsection{Morphologies of SNRs}\label{subsec:results_hd_mhd}

The HD evolution of SNRs in a uniform medium has been extensively explored \citep[e.g.][]{Cioffi1988,Iffrig2014, Li2015,Kim2015}. Analytical and numerical models can accurately predict the evolution and properties of the SNR during the different phases of its expansion. We first perform a qualitative and comparative analysis of the SNRs in a uniform medium at $0.25$~Myr in Fig.~\ref{fig:uniform_slices}. The first row shows thin slices of the HD simulation with $n=10$~cm$^{-3}$ across the $XZ$ plane that cuts through the centre of the SNR, showing the gas density and temperature. By using equation 7 in \cite{Kim2015}, we estimate the formation of the dense, radiative shell to occur at $t_\text{\rm dsf} \approx 0.012$~Myr. The presented SNR at $0.25$~Myr is thus depicted well beyond the end of the Sedov-Taylor phase. Nevertheless, the well-known features of a SNR can be easily distinguished even at this late time: a dense shell formed behind the forward shock is followed by a low density and high temperature bubble. The outer edge of the shell shows a slight increase in temperature, due to shock heating of the ambient gas. However, the density of the shell at this time is shaped by efficient radiative cooling, resulting in about 4 orders of magnitude lower temperature than the low density bubble. Furthermore, the shape of the shell has diverged from its spherical configuration, now exhibiting multiple irregularities. At this stage of the SNR evolution, the shell has become thick enough so that the thin shell approximation of the Sedov-Taylor solution is no longer valid. The formation of the shell is followed by the emergence of a radiative reverse shock, which results in a shell that is bounded by shocks on both sides. This results in the growth of shell instabilities \citep[e.g][]{Vishniac1983, Blondin1998, Kim2015}, driven by simulating the radial expansion of the gas on a Cartesian grid. The size of these perturbations reaches $\sim 10$\% of the SNR radius, in agreement with 2D simulations of \cite{Blondin1998} and the 3D simulations of \cite{Kim2015}.

In the second row of Fig.~\ref{fig:uniform_slices} we include the equivalent density and temperature slices for the MHD simulation with $B_0 = 1\,\mu$G, including an additional slice for the ratio of the thermal ($P_{\text{th}}$) to total pressure which in general is composed of the following terms:
\begin{align}\label{eq:total_pressure}
    P_{\text{tot}} = P_{\text{th}} + P_{\text{mag}} + P_{\text{ram}} + P_{\text{CR}},
\end{align}
where $P_{\text{mag}} = \vert \vec{B} \vert ^2/(8\pi)$ is the magnetic pressure and $P_{\text{ram}} = \rho \vert \vec{v} \vert ^2$ is the ram pressure, with $\vec{v}$ the gas velocity vector. The ambient medium in this simulation is magnetically dominated (see Table~\ref{tab:init_cond_ambient}). The immediate effect of adding magnetic fields is a morphological difference with respect to the HD simulation: the shell is wider in the equatorial direction (perpendicular to the initial magnetic field) than at the poles (parallel to the initial magnetic field). We find this effect to be more pronounced at later times. The black coloured streamlines overlaid on the density slice show the magnetic field direction $\vec{B}$ in the plane. Field lines are bent by the shock front of the shell, as expected for the frozen-in condition of ideal MHD. The addition of significant magnetic pressure explains why the shell has a lower $P_{\text{th}}/P_{\text{tot}}$. In turn, this additional pressure suppresses instabilities in the shell when compared to the HD case as the bending of magnetic field lines is opposing gas fragmentation. It is only as we approach the inner edge of the shell that $P_{\text{th}}$ becomes dominant. The shell departs from the relatively thin and condensed shell of the HD simulation in the equatorial regions. The faster travelling forward and reverse shocks allow for the magnetic pressure to further widen the shell as it moves to the low density bubble. This widening of the shell results in a hot bubble with significantly lower volume ($\geq 25$\% at this time, but it reaches 3 times lower volume at $1$~Myr) than the HD simulation and slightly higher ($\sim 33$\%) temperature (see Section~\ref{subsec:profiles_hd_mhd} for a more complete discussion). The shell-widening effect on the volume of the bubble provides a mean for re-compression of the hot gas at these late times, once the magnetic energy of the shell becomes higher than its thermal energy. These morphological differences with HD simulations are in good agreement with previous work on individual SNe in a uniform magnetised medium \citep[e.g.][]{Ferriere1991a, Hanayama2006, Hennebelle2014, Kim2015}.

The last row of Fig.~\ref{fig:uniform_slices} shows equivalent slices for the CRMHD simulation with an initial fraction of CR energy $\chi_{\text{CR}}=0.05$. Similarly to the MHD case, an initially uniform magnetic field introduces an asymmetry, which is now even more pronounced due to the CR pressure. Furthermore, the CRMHD simulation displays new features at the SNR poles, not present in the HD and MHD cases. Specifically, there is a strong conical disturbance dominated by the CR pressure (see last column) which also exhibits higher ram and magnetic pressure. These enhanced pressures push the high density shell further along the $z$ axis with the entire region of the shell in the inner $\sim 5$~pc around the $z$ axis displaying an outward blowout. Such `inflating' blisters have been observed in SNRs, like Vela \citep{Meaburn1988}, and attributed to the rupture of the shell by the growth of Vishniac instabilities \citep{Pittard2013}. The gas density drops directly above the poles of the remnant. The anisotropic morphology of this CRMHD run resembles the findings by \cite{Girichidis2014} and \cite{Dubois2016} and we will examine its origin in more detail in the next section.

\subsection{Radial structure of SNRs}\label{subsec:profiles_hd_mhd}

\begin{figure*}
  \centering \includegraphics[width=\textwidth]{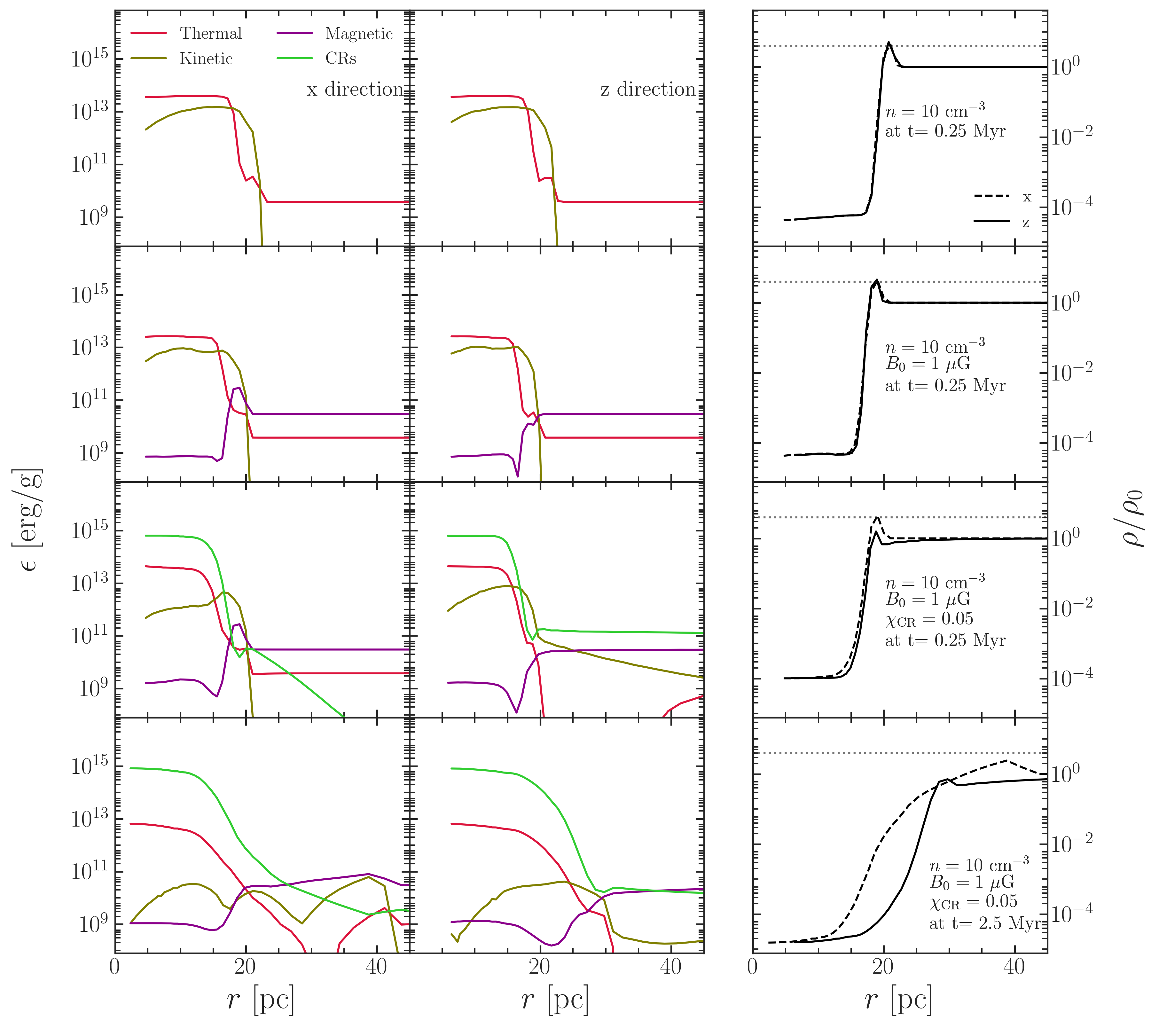}
  \caption{Comparison of radial specific energy and density profiles for the HD, MHD and CRMHD simulations with $n= 10$ cm$^{-3}$, initially uniform magnetic field of 1~$\mu$G and $\chi_{CR} = 0.05$. The first two columns show the energy profiles along the $x$ and $z$ directions, respectively. The rightmost column shows the density profiles along the $x$ (dashed line) and $z$ (solid line) directions. Line colours in the energy profiles indicate different components (see legend at the top left). All simulations are compared at $0.25$~Myr in the first three rows. We also include profiles at $2.5$~Myr for the CRMHD run (bottom row). For the MHD run, the specific magnetic energy profile along the $z$ axis shows a decrease with respect to the ambient medium, both in the shell and in the bubble, while in the $x$ direction there is an enhancement of the specific magnetic energy within the dense shell. CRMHD profiles in the $z$ direction show a smaller SNR, smoother thermal profiles and an increased kinetic energy density of the ambient gas. The contribution of CRs to the energy density in the bubble exceeds the other energies in both $x$ and $z$ directions. At 2.5 Myr, the density profiles for the CRMHD run show that the SNR has grown from $\sim$19~pc to $\sim$39~pc in the $x$ direction with an overall decrease of the thermal and kinetic energies of the shell and bubble.}
  \label{fig:density_energy_profile_comp}
\end{figure*}

We now move beyond the qualitative analysis to study the radial structure of the SNR and its time evolution. In particular, understanding the distribution of energies and densities across the shock provides us with a valuable insight into the physics driving the deposition of SN momentum. To this end, Fig.~\ref{fig:density_energy_profile_comp} compares our fiducial set of HD, MHD and CRMHD simulations (i.e. with $n=10$ cm$^{-3}$, $B_0 =1$~$\mu$G and $\chi_{\text{CR}}=0.05$) at $t= 0.25$~Myr, where mass-weighted average radial profiles of specific energies (left and middle columns) together with radial gas density profiles (rightmost column) are shown.

The Rankine-Hugoniot relations for a strong adiabatic shock of an ideal gas with $\gamma =5/3$ predict $\rho_2/\rho_1 \approx 4$, with $\rho_2$ and $\rho_1$ being the post-shock density and the pre-shock density, respectively. In the rightmost panel of the first row of Fig.~\ref{fig:density_energy_profile_comp}, the HD simulation has a jump in the gas density of $\gtrsim 4$, due to the fact that at 0.25 Myr the shell has become radiative, and a higher compression can be reached. The position of the density jump coincides with the jump of gas kinetic energy (i.e. the shell), followed by a jump of about 4 orders of magnitude in thermal energy. This high thermal energy region spans the same range as the drop in the gas density, and it corresponds to the inner hot bubble. These profiles provide a shell radius estimate of $\sim 19$~pc at $0.25$~Myr for this HD run.

The MHD case is shown in the second row of Fig.~\ref{fig:density_energy_profile_comp} confirming the visual impression shown by Fig.~\ref{fig:uniform_slices}; i.e. the presence of an initially aligned magnetic field in the medium where the SN takes place generates an expanding SNR with a clear anisotropy. Magnetic energy is amplified in the $x$ direction (perpendicular to the magnetic field), but not in the $z$ direction (parallel to the magnetic field). The magnetosonic shock travels faster in the $x$ and $y$ direction, while expansion in the $z$ direction is dictated by the sound speed as in the HD simulations. As the shock travels in the transverse direction, magnetic field lines are compressed, increasing the magnetic pressure behind the outer shell. This extra pressure provides the means for the reverse shock to compress the hot gas of the bubble further. These deviations from the HD case are already present at $0.25$~Myr and become more prominent with time. This supports the claim that magnetised media affect the evolution of SNRs when the magnetic pressure in the shell is comparable to the thermal pressure \citep[e.g.][]{Ferriere1991a,Iffrig2014,Kim2015}. The broadening of the shell along the $x$ and $y$ directions coincides with a region of enhanced magnetic energy. This is followed by another peak in the kinetic energy profile, corresponding to the reverse shock. Due to the additional magnetic energy in the $x$ direction, this inward shock carries higher momentum and has travelled further towards the centre of the SNR than in the HD case. This additional momentum, while not being important for the total radial momentum, could feed higher turbulence in the bubble.

\begin{figure}
  \centering \includegraphics[width=\columnwidth]{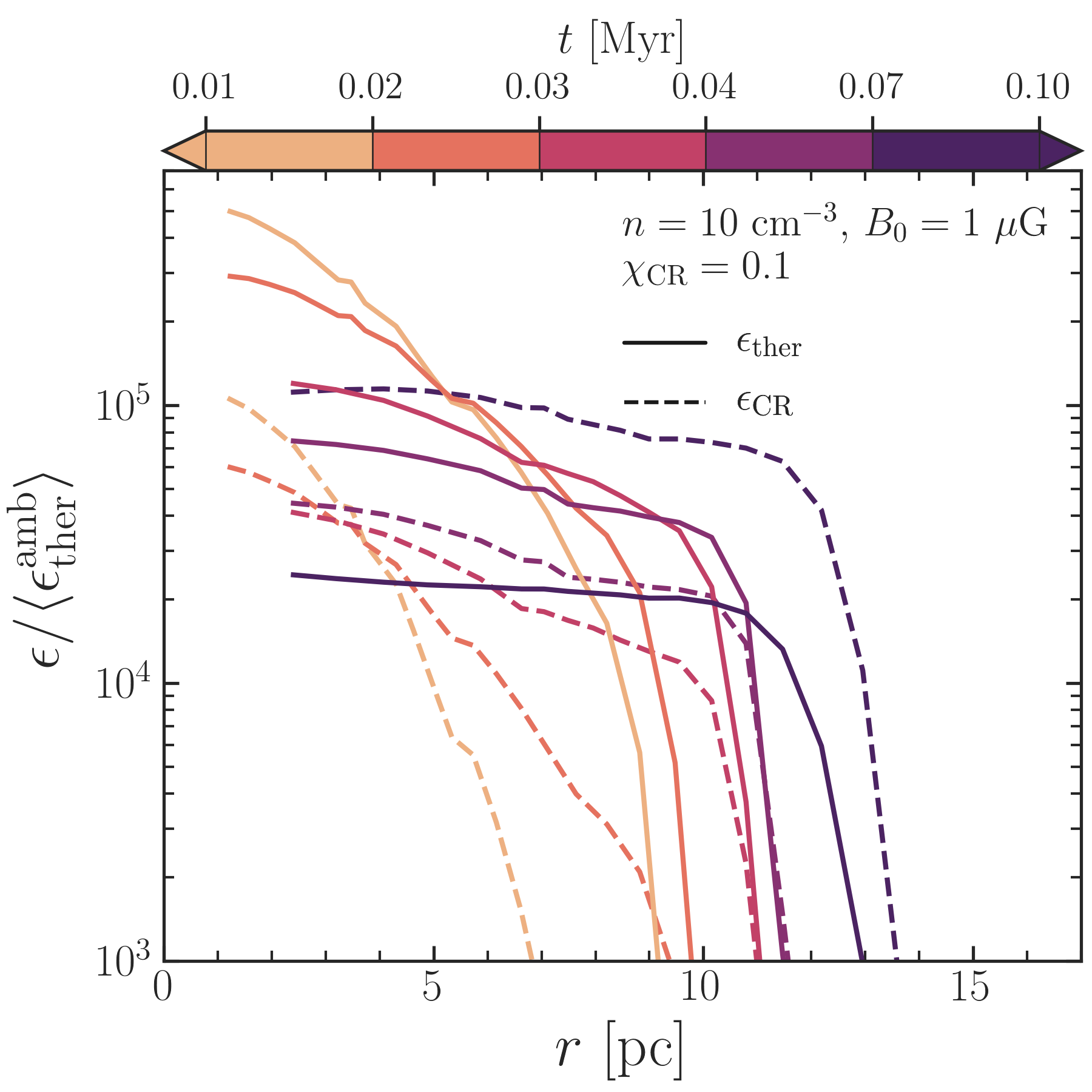}
  \caption{Time evolution of specific thermal (solid lines) and CR (dashed lines) energy profiles in the $x$ direction (i.e. perpendicular to direction of the initial magnetic field) for the CRMHD simulation with $n= 10$ cm$^{-3}$, $B_0 = 1\,\mu$G and $\chi_{\rm CR} = 0.1$. Profiles are normalised to the average specific thermal energy of the ambient medium $\langle\epsilon_{\text{ther}}^{\text{amb}}\rangle$ at each time. The solid lines show the decrease of the central thermal energy as the SNR  expands. Before $0.07$~Myr, thermal energy dominates over CR energy inside the bubble. However, at later times this power relation is swapped, with CR energy dominating over thermal energy both in the bubble and the shell.}
  \label{fig:thermal_evolution_zoom}
\end{figure}

We now turn to the analysis of our fiducial CRMHD run. The most significant feature is the fact that the CRs dominate the energy budget inside the SNR by $\sim 2$ orders of magnitude (recall that $\chi_{\text{CR}} = 0.05$ is the fractional energy contribution of cosmic rays at the start of the simulation). This is expected due to CRs expanding with a lower adiabatic index. The dominance of CRs inside the SNR increases with time as shown at $2.5$~Myr in the bottom row of Fig.~\ref{fig:density_energy_profile_comp}.

In the CRMHD run, the density profiles in the $x$ and $z$ direction (third row, rightmost panel of Fig.~\ref{fig:density_energy_profile_comp}) present a clear anisotropy, much more pronounced than for the MHD case. This anisotropic expansion is not limited to the gas properties alone. As explained in Section~\ref{sec:introduction}, the diffusion and streaming of CRs is mediated by magnetic field lines, which means that we expect a more significant flow of CRs through the poles ($z$ direction) than through the equatorial region, as shown in Fig.~\ref{fig:uniform_slices}. While the density profile in the $x$ direction resembles the MHD case, in the $z$ direction it shows a much lower shock discontinuity which is also lagging $\sim 1$ pc behind the MHD case. Diffusion of CRs creates an extended pressure gradient that increases the kinetic energy of the ambient gas and disrupts the morphology of the shell. This effect was previously observed by \cite{Dubois2019} in the context of a simulated turbulent and inhomogeneous ISM. This effect becomes stronger with time such that at $2.5$~Myr (Fig.~\ref{fig:density_energy_profile_comp}, bottom row) the shock in the $z$ direction has become a smooth bump in the density profile. Therefore momentum is imparted to the ambient medium by the expanding shell as well as by the CRs diffusing along the poles. The pressure exerted by the escaping CRs is not just affecting the structure of the shock, but is also reducing the gas density ahead of it. Thus, it becomes more difficult for the shell to accumulate gas. 

In order to study further the impact of CRs, we compare in Fig.~\ref{fig:thermal_evolution_zoom} the thermal (solid lines) and CR specific energy (dashed lines) profiles for the CRMHD run, for our run with the highest explored value of $\chi_{\text{CR}} = 0.1$. Each profile is normalised to the average specific thermal energy of the ambient medium $\langle\epsilon_{\text{ther}}^{\text{amb}}\rangle$ at that time and the curves are colour-coded according to the simulation time. Both thermal and CR energy profiles show how, as the bubble volume increases with time, the specific energy decreases due to adiabatic expansion. However, while the thermal energy profile keeps decreasing as time progresses, the CR energy profile appears to stall, reaching equipartition at $\sim 0.04-0.07$~Myr. After $0.07$~Myr, CRs begin to dominate the energy budget of the bubble and the shell. We will review this important change in dynamics in Section~\ref{subsec:momentum_hd_mhd}.

\subsection{Time evolution of the SNR momentum deposition}\label{subsec:momentum_hd_mhd}
\begin{figure}
  \centering \includegraphics[width=\columnwidth]{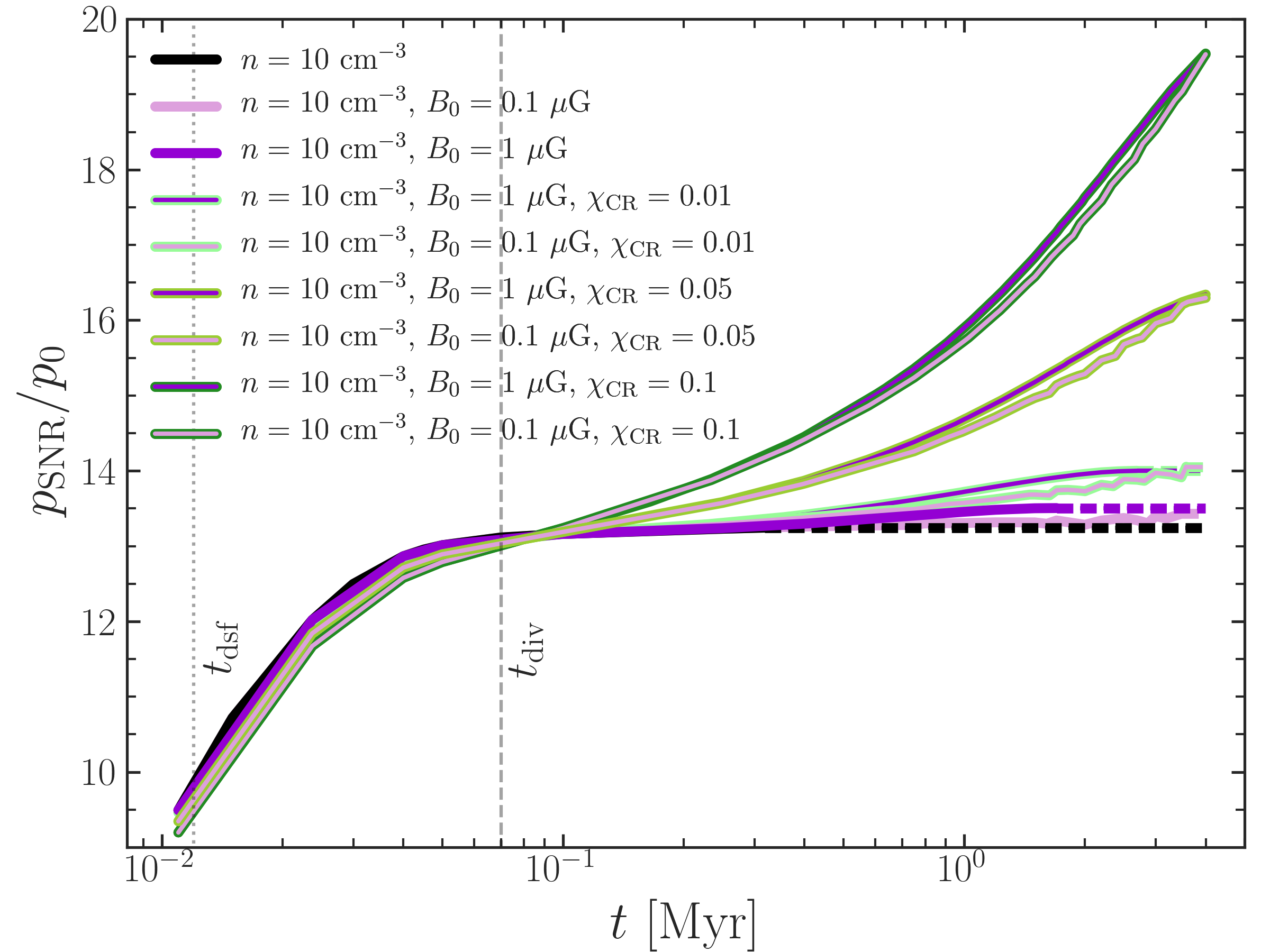}
  \caption{Evolution of the total radial momentum of the SNR $p$ with time for the simulations with $n=10$~cm$^{-3}$, $B_0 = 0$, 0.1 and 1~$\mu$G and $\chi_{\text{CR}} = 0$, 0.01, 0.05 and 0.1. For CRMHD simulations the inner line colour indicates the strength of the initial magnetic field in purple shades (darker shades indicating stronger magnetic fields) and the outer contour colour indicates the value of $\chi_{\text{CR}}$ in green shades (darker shades corresponding to higher $\chi_{\text{CR}}$). After $\sim 0.1$~Myr, the evolution of $p_{\rm SNR}$ strongly depends on the contribution of CRs to the energy budget of the SNe (e.g. $\sim 50\%$ increase with respect to the HD simulation for M1HD10CR10). Once shell fragmentation due to instabilities begins, we extrapolate the final momentum, represented by dashed lines. The vertical dashed line indicates the point where CRMHD runs surpass in momentum the HD run, $t_{\rm div}$, and the doted line is $t_{\rm dsf}$ from equation 7 of \protect\cite{Kim2015}. A higher magnetic field and/or $\chi_{\text{CR}}$ delays fragmentation.}
  \label{fig:momentum_vs_t}
\end{figure}

We now turn to study the total radial momentum of SNRs and compare our findings with the classical understanding of SNR expansion in a uniform medium. In Fig.~\ref{fig:momentum_vs_t} we show the time evolution of radial linear momentum of the shell and bubble combined, measured as $\vec{p}_{\text{SNR}} = \vec{p}_{\text{out}} + \vec{p}_{\text{in}}$ ($\vec{p}_{\text{out}}$ and $\vec{p}_{\text{in}}$ are the outwards and inward radial momentum vectors of the gas, respectively) for our set of fiducial simulations. We show the evolution from $t \geq 0.01$~Myr, shortly before the estimated radiative, dense shell formation time $t_{\rm dsf}$ \cite[their equation 7]{Kim2015}. $p_{\text{SNR}}$ is normalised by the initial momentum $p_0=\sqrt{2m_{\rm ej}E_{\rm SNe}}$ for a single SN of $E_{\rm SNe}=10^{51}$~erg and ejecta mass of $m_{\rm ej}=2M_{\mathrm \odot}$, following \cite{Naab2016}. The value of $t_{\rm dsf}$ is the time at which cooling effects of the gas become non-negligible, and is a good estimate for the end of the Sedov-Taylor phase \citep{Kim2015}. The black solid line shows the data for the HD simulation with $n=10$~cm$^{-3}$. For $t > 0.04$~Myr momentum increases mildly, due to the start of the adiabatic pressure-driven snowplough, stabilising to its final value at $\sim 0.5$ Myr in the momentum-conserving phase.

We now turn to examine the momentum deposited in the MHD runs, which displayed an axisymmetric expansion of the shell. In Fig.~\ref{fig:momentum_vs_t} we show $p_{\text{SNR}}$ for the MHD simulations with ambient density $n=10$~cm$^{-3}$ and two initial magnetic field strengths: $B_0=0.1, 1\;\mu$G. Both MHD simulations follow closely the momentum gain of the HD simulation, with only $\sim 3-4$\% higher momentum at $4$~Myr. These results indicate that the addition of an initially uniform magnetic field does not strongly influence the final radial momentum of a SNR, in agreement with previous work \citep[e.g.][]{Iffrig2014,Kim2015}.

We contrast these results with our findings for the CRMHD runs, where we vary both magnetic field strength and the fraction of SN energy that goes into CRs, $\chi_{\text{CR}}$. Before the inflection point at around $0.07$~Myr, CRMHD runs have midly lower momentum than their HD and MHD analogues. Since CRs have a lower adiabatic index, the Sedov-Taylor phase with CRs will always have a lower momentum than an ideal gas without CRs \citep{Sedov1959}. At early times, during the adiabatic phase, and shortly after, the self-similarity of the Sedov-Taylor solution is conserved, in agreement with analytical studies \cite{Chevalier1983}. 
 
After $\sim0.08$~Myr, we find the CRMHD simulations to experience a new phase of renewed momentum growth previously unobserved in numerical simulations, predicted by the semi-analytical work by \citet{Diesing2018}. After this period, CRMHD runs diverge from the snow-plough plateauing of the HD and MHD runs (see Section~\ref{subsec:results_modelling} for more details). At later stages, the simulations with CRs continue to gain momentum. This increase in the gained momentum is faster and larger for a higher $\chi_{\text{CR}}$ but largely independent from the magnetic field strength. We find 7\%, 27\% and 50\% more radial momentum than in the HD case at 4~Myr, for $\chi_{\text{CR}}=0.01$, 0.05 and 0.1, respectively.

\subsection{Modelling the momentum deposition of SNe with CRs}\label{subsec:results_modelling}

\begin{figure*}
  \centering \includegraphics[width=\textwidth]{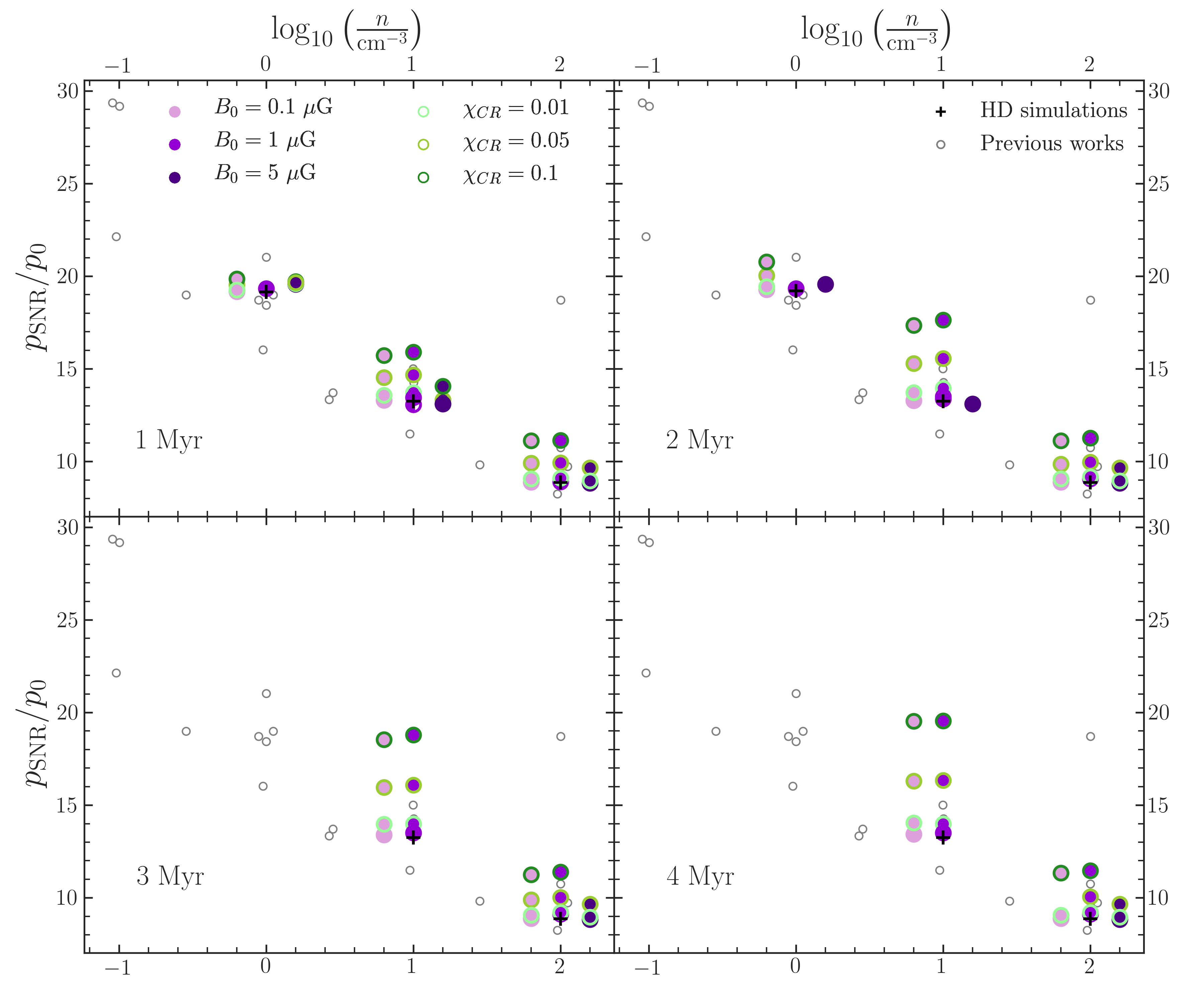}
  \caption{Total radial momentum of the SNR (i.e. bubble and shell combined, with $p_{\text{SNR}}=p_{\text{out}}-p_{\text{in}}$), as a function of the ambient number density $n$ for HD (crosses), initially uniform MHD (circles) and CRMHD (green outlined circles) simulations. Each of the four panels shows a different time (i.e. 1, 2, 3 and $4$~Myr from top left to bottom right, respectively). Simulations that did not reach the 3 and 4~Myr marks are excluded from the corresponding panels. Purple colours used for the circles filling indicate the initial magnetic field strength. The green outline of each circle indicates the value of $\chi_{\rm CR}$ for its corresponding CR run. $p_{\text{SNR}}$ is normalised by the \protect\cite{Kim2015} momentum $p_0=14181 M_{\odot}$~km~s$^{-1}$ for $E_{\rm SN} = 10^{51}$ erg. Empty grey circles correspond to previous works compiled in figure 5 of \protect\cite{Naab2016} (see references therein). Our markers are manually displaced by a small amount along the $x$-axis depending on their magnetic field strength to facilitate reading. Our HD simulations lie close to the results of previous works, not showing any significant evolution from 1 to $4$~Myr. While MHD and CRMHD runs have similar values to the HD runs at $1$~Myr, they exhibit a significant evolution across the four panels. MHD runs gain an additional 8-10\% momentum with respect to their HD counterparts at $n=10$~cm$^{-3}$ and $\sim 5$\% at $n=100$~cm$^{-3}$. This increase is significant for 1~$\mu$G, but not for 0.1 or 5~$\mu$G. However, the introduction of CRs increases the deposited momentum across all densities and initial magnetic fields, with higher $\chi_{\rm CR}$ monotonically yielding a higher momentum increase. This difference in the CRMHD runs is time dependent, reaching a value of $\sim50$\% with respect to their MHD counterpart.}
  \label{fig:final_momentum_comparison}
\end{figure*}

Fig.~\ref{fig:momentum_vs_t} provided us with evidence that the SNR momentum deposition has some dependence on the fraction of SN energy initially contained in CRs. To quantify this result using our entire simulation suite (all in Table~\ref{tab:sims}, except runs with tangled magnetic field ICs) we show in Fig.~\ref{fig:final_momentum_comparison} the total radial linear momentum as a function of the ambient number density, providing an update of \cite{Naab2016} (see their figure 5). Different panels show the value of $p_{\rm SNR}/p_0$ at 1, 2, 3 and 4~Myr. Simulations with low ambient density ($n= 1$~cm$^{-3}$) are not included after 2~Myr since our computational domain is not large enough to capture their evolution.

Focusing first on the HD simulations (black crosses), we find they are in good agreement with previous numerical studies (open grey circles) \cite[data compiled by][]{Naab2016}. The final momentum decreases with increasing ambient density, according to the Sedov-Taylor solution \citep{Sedov1959} and in agreement with previous studies \cite[e.g.][]{Cioffi1988,Kim2015,Iffrig2014,Martizzi2015}.
The increase of momentum from $t_{\rm dsf}$ to 1~Myr is about $\sim 40$\%, in agreement with the final momentum deposition increase of 50\% found by \cite{Kim2015}. The momentum at $t_{\rm dsf}$, $p_{\text{dsf}}$, \citep[their equation 7][]{Kim2015} and the momentum at 1~Myr $p(t=1 \text{ Myr})$ as measured in our simulations can be fitted by the following power-laws:
\begin{align}
    &p_{\text{dsf}} = 1.94\times10^5 M_{\odot}\text{ km s}^{-1} \left(\frac{n}{\text{cm}^{-3}}\right)^{-0.13} \\
    &p(t=1 \text{ Myr}) = 2.68\times10^5 M_{\odot}\text{ km s}^{-1} \left(\frac{n}{\text{cm}^{-3}}\right)^{-0.16}\,.
\end{align}
The momentum at $1$~Myr, while having a slightly lower dependence on the ambient number density $n$, is consistent with the results of \cite{Kim2015} and \cite{Iffrig2014}.

As discussed in Section~\ref{subsec:momentum_hd_mhd}, the effect of an initially aligned magnetic field on the radial momentum of the SNR is negligible. This is confirmed for all the ambient number densities explored in this work, and illustrated in Fig.~\ref{fig:final_momentum_comparison} by the circles filled with purple colours, where the momentum gain with respect to the HD runs is within 10\%. However, in the presence of CRs the total momentum changes noticeably, especially at later times. The data points diverge accordingly from the locus in the $p_{\rm SNR}/p_0-n$ plane occupied in the absence of CRs. This departure is 
most significant for the largest values of $\chi_{\rm CR}$ and intermediate ambient number densities we explored. However, it seems unaffected by the strength of the initial magnetic field.  

Based on our CRMHD simulations, we propose an updated version of the deposited momentum law that accounts for the dependence on the CR contribution to the energy of the SNe:
\begin{align}\label{eq:crmhd_mom_fit}
    p_{\rm SNR} (n, \chi_{\text{CR}}) = A (\chi_{\text{CR}} + 1)^{\gamma}\left(\frac{n}{\text{cm}^{-3}}\right)^{\alpha},
\end{align}
where $A$, $\gamma$ and $\alpha$ are parameters obtained by fitting our results. However, when exactly to measure the deposited momentum is not so clear for the CRMHD runs. For the well-studied HD case, this is done when the remnant reaches the momentum conserving phase, as it has been done in multiple analytical and numerical studies \citep[e.g.][]{Cioffi1988,Thornton1998,Hennebelle2014,Kim2015,Li2015}. Figs.~\ref{fig:momentum_vs_t} and \ref{fig:final_momentum_comparison} show that the CRMHD runs with density 10 cm$^{-3}$ do not show an unequivocal flattening trend, except for the runs with $\chi_{\text{CR}}=0.01$, for which $p_{\rm SNR}/p_0$ appears not to grow by more than a few percent after $\sim 2$~Myr. This lack of plateauing in the radial momentum is not ubiquitous across our CRMHD runs: for $n=100$~cm$^{-3}$ the final momentum is well approximated by its value already at around 1~Myr, as shown by Fig.~\ref{fig:final_momentum_comparison}. As discussed in Section~\ref{subsec:momentum_hd_mhd}, the additional momentum growth for the CRMHD runs is due to the change of energy dominance inside the bubble of the SNR, from thermally to CR dominated. Since at the same density and magnetic field, runs with higher $\chi_{\text{CR}}$ conserve a higher CR energy budget inside the bubble, this turnover occurs earlier with a higher CR fraction. On the other hand, previous works showed that for HD simulations \citep[e.g][]{Naab2016} single SN in a higher density ambient medium experiences a higher radiative cooling. This means that for higher densities, SNRs become CR dominated earlier. Thus, we expect that the growth in momentum observed at lower densities will eventually be halted and lead to momentum flattening in the same manner as we found for the high density runs.
\begin{table}
 \caption{Fitting parameters for the final radial momentum $p_{\rm SNR}$ power laws. The parameters are described by Eq.~\ref{eq:crmhd_mom_fit}. For direct comparison we also provide the results for the HD simulations.}
 \label{tab:fitting_results}
 \centering
 \begin{tabular}{lccc}
  \hline
  Run type & $A$ [$M_{\odot}$ km s$^{-1}$] & $\gamma$ & $\alpha$\\
  \hline
  HD & $2.68\times 10^{5}$ & – & -0.160\\
  CRMHD & $2.87\times 10^{5}$ & 4.82 & -0.196\\
  \hline
 \end{tabular}
 \end{table}

Our model is based on the ansatz that SNRs with CRs experience an additional evolutionary phase. Thus, to model the momentum evolution, the first step is finding the moment when the momentum of the CRMHD run overcomes that of its corresponding HD run at the same ambient density. As an example, the start of this phase for the runs with $n=10$~cm$^{-3}$ appears at $\sim 0.09$ Myr (see dashed vertical line in Fig.~\ref{fig:momentum_vs_t}), with a small dependence on $\chi_{\text{CR}}$. Following this, our model finds the point of inflection where the curvature changes to concave-down. Since the momentum evolution is a smooth function of time, we apply a cubic spline interpolation for a better estimation. We consider this as the point at which the acceleration due to the CR energy density starts to decrease, which would in turn be followed by the turnover of the momentum gain towards the final momentum in this final \textit{CR pressure snowplough} phase. Finally, after finding the inflection point where the curvature changes to concave down, we look for the maximum momentum achieved in the CRMHD runs that have converged. If no final convergence has been captured but the final inflection point has been reached, we estimate the expected maximum momentum by assuming no further acceleration occurs. In Appendix~\ref{ap:example_momentum_model} and Fig. \ref{fig:model}, we present examples of the computation for the final momentum when convergence is not being captured, and the final results are given in Table~\ref{tab:fitting_results}. Contrary to the results obtained by \cite{Diesing2018} in their semi-analytical modelling of SNRs with CRs, our findings have a stronger boost for lower ambient densities. Additionally, we find an overall smaller boost: e.g. for $n=1$~cm$^{-3}$ and $\chi_{\rm CR}=0.1$ we obtain a boost of a factor of $\sim 1.7$, while their model is closer to 2-3.

\section{Results with Tangled Magnetic Fields}\label{sec:results_tangled}
In previous sections we have explored the evolution of individual SNe in a uniform medium with an initially uniform magnetic field aligned along the $z$ axis. This configuration provided us with a simple starting point where the influence of different physical components, such as thermal, magnetic and CR energies, can be distinguished. However, this setup can be improved to provide a better approximation to the realistic evolution of a SNR in the ISM \citep[e.g.][]{Iffrig2014, Walch2015,Geen2016}.
The large-scale distribution of the Galactic magnetic field (GMF) has been extensively explored through rotation measures and polarised synchrotron emission \citep[e.g.][]{Sun2008,Jansson2012,Polderman2020}. Coherent magnetic fields may lead to e.g. a bipolar morphology for the radio emission \citep{West2015}. However, observations find turbulent magnetic fields to dominate the magnetic energy budget \citep{Beck2007,Beck2015}. These turbulent, small-scale magnetic fields may even dominate the evolution of young remnants \citep{Rand1989,Ohno1993}. 

The magnetic field configuration is crucial for the propagation of CRs. While our uniform magnetic field leads to anisotropic diffusion along the magnetic field initial direction, the turbulent magnetic fields in the ISM provide a more isotropic evolution of CRs. Therefore, in this section we explore an initial configuration of the magnetic field that better represents magnetic fields in the ISM. Our magnetic ICs follow a Kazantsev energy spectrum, characteristic of turbulent dynamo amplification in the ISM of galaxies \citep{Schekochihin2002,Federrath2014}. This inverse-cascade spectrum concentrates magnetic power at small scales and provides an effective diffusion coefficient one third lower than the maximum value in the aligned case, $D_0$.

\subsection{Morphologies of SNRs with tangled magnetic fields}\label{subsec:topology_tangled}
\begin{figure*}
  \centering \includegraphics[width=\textwidth]{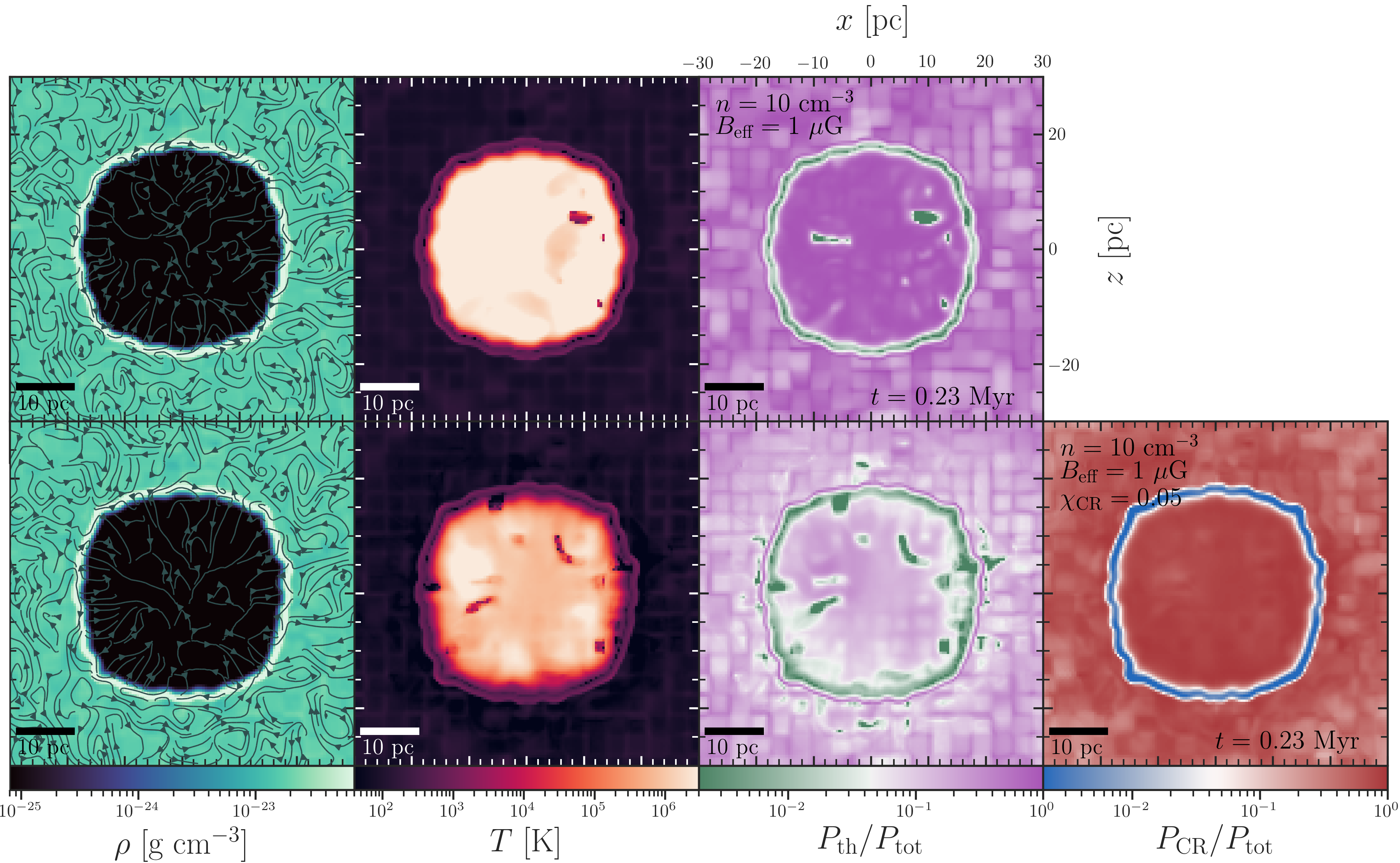}
  \caption{Slices in the $XZ$ plane for the MHD (top row) and CRMHD (bottom row) SN simulations with magnetic initial conditions. The ambient density is initialised to $n=10$~cm$^{-3}$, $B_{\rm eff} = 1$~$\mu$G, and $\chi_{\rm CR}=0.0$ (top row), and $0.05$ (bottom row). Slices are shown $0.23$~Myr after the start of the simulation. \textit{First column}: gas density with overlaid streamlines showcasing the magnetic field lines projected onto the $XZ$ slice. \textit{Second column}: gas temperature. \textit{Third column}: thermal to total pressure ratio. \textit{Fourth column}: CR to total pressure ratio. The isotropic distribution of the magnetic field in the ambient medium at large scales generates a spherical SNR, resembling the result of the HD simulation. There is a clear separation between a dense, highly magnetised shell and a low-density, hot bubble. Within the hot bubble there are some magnetically-dominated patches with lower temperatures. In contrast to the strongly anisotropic diffusion of CRs in the aligned CRMHD runs, the CR pressure is now isotropic and dominates both in the bubble and the surrounding medium.}
  \label{fig:slices_magics}
\end{figure*}

Before turning to a quantitative analysis of how this tangled magnetic configuration affects the momentum deposition of the SNR, we perform a brief qualitative review of the morphology of the resulting SNRs with $B_{\rm eff} = 1$~$\mu$G. In the top row of Fig.~\ref{fig:slices_magics} we present slices of these MHD runs across the $XZ$ plane. These are shown at $\sim 0.23$ Myr, with the colourmap limits as in Fig.~\ref{fig:uniform_slices}. The gas density slice evidences that the asymmetry observed in the aligned MHD runs vanishes. In fact, the morphology of the SNR resembles the HD case (see top row of Fig.~\ref{fig:uniform_slices}). The projected magnetic field lines in the ambient medium clearly show twist and bends indicative of their tangled nature.

As we have specified in Section~\ref{sec:methods_ics}, our tangled magnetic field configuration is setup in an ambient medium of uniform gas density. Such a configuration leads to the emergence of magnetically-driven perturbations, as shown by the middle and rightmost panels in the first row of Fig.~\ref{fig:slices_magics}. These panels show the gas temperature and the ratio of the thermal ($P_{\text{th}}$) to the total pressure ($P_{\text{tot}}$), respectively. The new ambient medium is filled with patches where either the thermal pressure (purple) or the magnetic + ram pressure (pink/white) dominate. We also find regions of few parsecs within the bubble where the thermal energy is subdominant. The shell thickness is also reduced compared to the aligned MHD case, and consequently thin shell instabilities are triggered, similar to the case of the HD simulations. The angle of the magnetic field lines to the shock normal changes along the shell perimeter and the exact dependence of CR acceleration efficiency on this angle remains an open question \citep{Pais2018a,Dubois2019}. Regardless of any differences in the appearance of the SNR, in Section~\ref{subsection:momentum_tangled} we show that our tangled magnetic field setup does not affect the momentum deposition of the SNR, which is the main focus of this paper.

The volume of the hot bubble in the SNR simulated with a uniform magnetic field is maximal at $\sim 1$ Myr. Afterwards, this volume decreases as the horizontal compression due to magnetic pressure overcomes the volume increase due to the SNR expansion. This reduction in the volume fraction of hot gas agrees with previous works \citep[e.g.][]{Hanayama2006, Hennebelle2014, Kim2015}. In contrast, the hot bubble in the run with tangled magnetic fields follows a similar evolution to the HD run. The volume of this bubble is in fact almost 6 times larger at $2$~Myr than the peak volume of the hot bubble in the aligned case. 

Having reviewed the general effects that a tangled magnetic field configuration has on our SNR in the absence of CRs, we now move to consider the CRMHD case. In the bottom row of Fig.~\ref{fig:slices_magics} we show slices of the CRMHD run with $n=10$~cm$^{-3}$, $B_{\rm eff}=1$~$\mu$G and $\chi_{\text{CR}}=0.05$, at $0.25$~Myr. CRs do not affect the tangled topology of magnetic fields noticeably and as in the case without CRs there is no significant magnetic shell widening. There are, however, some deviations in the shape and position of the shell irregularities suggesting that these disturbances are affected by the presence of CRs. Similarly to the case of aligned magnetic fields, magnetic and ram pressure dominate in the shell. CR pressure extends beyond the SNR and a halo of high CR pressure surrounds the shell.

\subsection{Momentum deposition of SNRs with tangled magnetic fields}\label{subsection:momentum_tangled}

\begin{figure}
  \centering \includegraphics[width=\columnwidth]{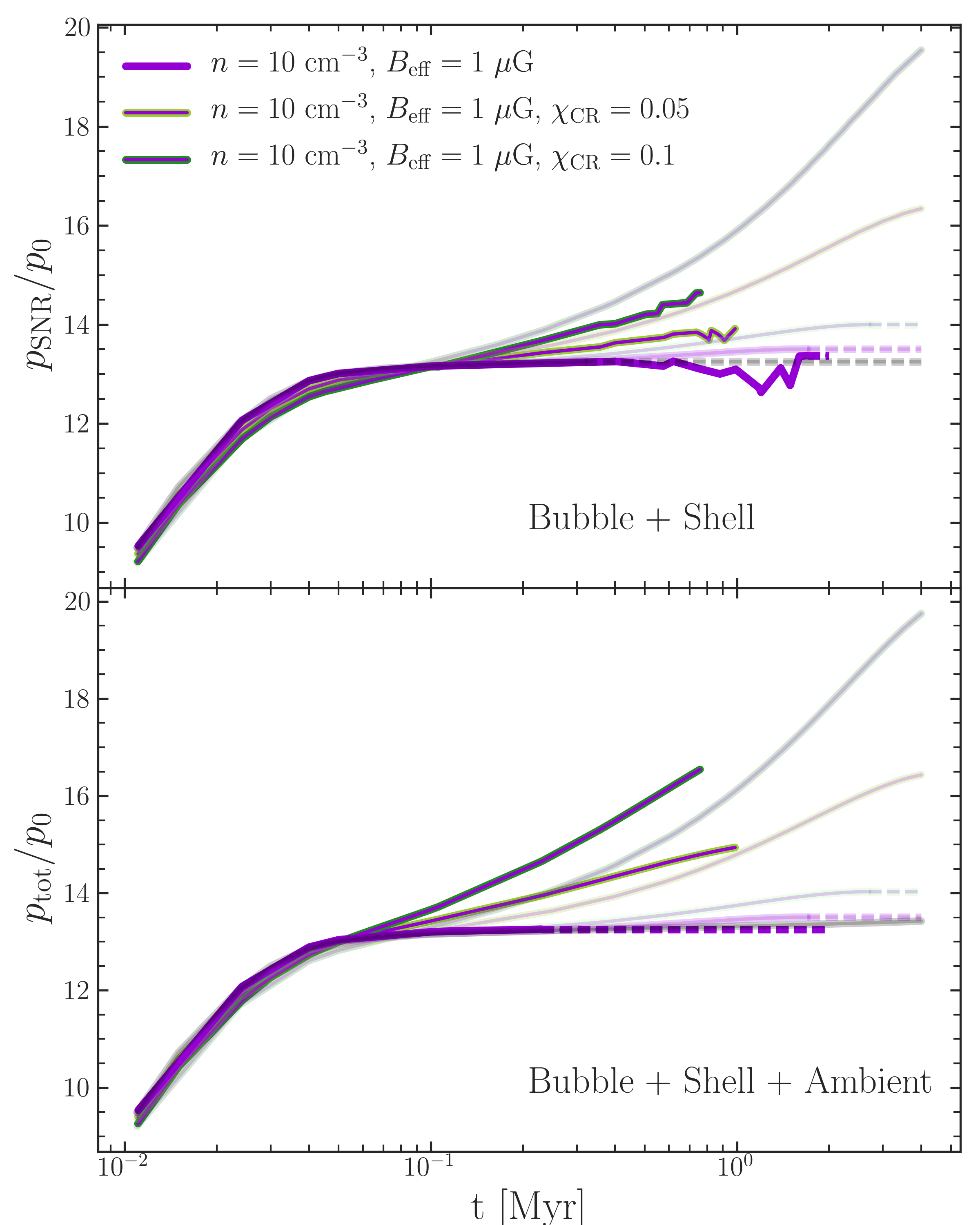}
  \caption{\textit{Top}: Time evolution of total radial momentum of the SNR in the MHD and CRMHD simulations with a tangled initial configuration of the magnetic field. We show runs with $n=10$~cm$^{-3}$, $B_\text{eff} = 1\mu$G and $\chi_{\text{CR}}=0.0$, 0.05 and 0.1. For comparison, we include in the background lines corresponding to the uniform magnetic field runs with equal magnetic field strength and ambient density. These are shown as lower opacity lines following the colour legend of Fig.~\ref{fig:momentum_vs_t}. \textit{Bottom}: Same as the top panel, now including the gas in the whole simulated box. The transition from solid to dashed lines indicates the onset of shell fragmentation. Fragmentation takes place earlier in the HD and tangled MHD runs. The SNR momentum gain of tangled runs falls below the matching uniform magnetic field cases when only accounting for the contributions from the bubble and shock shell. However, when the momentum of the ambient medium is included in $p_{\text{tot}}$, the momentum gain in tangled magnetic initial conditions runs exceeds the uniform magnetic field runs by $\sim 10$\%.}
  \label{fig:momentum_vs_t_tangled}
\end{figure}

In this section we examine the effect of a tangled magnetic field configuration on the momentum deposition of SNRs in our MHD and CRMHD simulations. We employ our region selection code (see Appendix~\ref{ap:region_algorithm} for further details) in the following analysis to better separate the non-homogeneous ambient medium from the SNR itself. Fig.~\ref{fig:momentum_vs_t_tangled} shows the evolution of the total radial momentum of the SNR (bubble and shell; top panel) and of the whole simulated box (bottom panel) for simulations with $n=10$~cm$^{-3}$, $B_\text{eff} = 1 \mu$~G, and $\chi_{\text{CR}} =0$, 0.05 and 0.1. To facilitate comparison we have included in this figure the matching HD, uniform MHD, and uniform CRMHD runs, which are shown by curves of same colour but with a lower opacity (legend as for Fig.~\ref{fig:momentum_vs_t}).

Firstly, we confirm again the result found with an aligned magnetic field configuration: the final momentum deposition is not affected by inclusion of magnetic fields, i.e. MHD runs both with aligned and tangled magnetic fields follow closely the evolution of the HD run. As for the uniform case, the final total radial momentum of the SNRs rises as we increase the fraction of initial SN energy deposited into CRs. When accounting solely for the momentum found in the bubble and shell, our runs with CRs and tangled magnetic fields have a systematically lower momentum gain than in the matching runs with uniform magnetic fields. Nonetheless, the total momentum deposited is higher in the tangled case when accounting for the momentum deposited in the immediate surroundings of the remnant (lower panel of Fig.~\ref{fig:momentum_vs_t_tangled}). This increase of about 10\% with respect to the uniform magnetic field runs is due to the acceleration provided by escaping CRs to the ambient gas beyond the SNR. While CRs impart more momentum to the shell in the case of uniform magnetic fields, in the tangled case the isotropic escape of CRs leads to a larger momentum transfer to the ambient gas. These runs have not reached momentum convergence by the final time of our simulations, and we expect additional momentum gains at later times. Therefore, it is worth stressing that all our simulations with CRs, irrespective of the magnetic field topology, have higher final momenta for the same initial SN energy than simulations without CRs.

\section{Mock observations of simulated SNRs}\label{sec:mocks}

\begin{figure*}
  \centering \includegraphics[width=\textwidth]{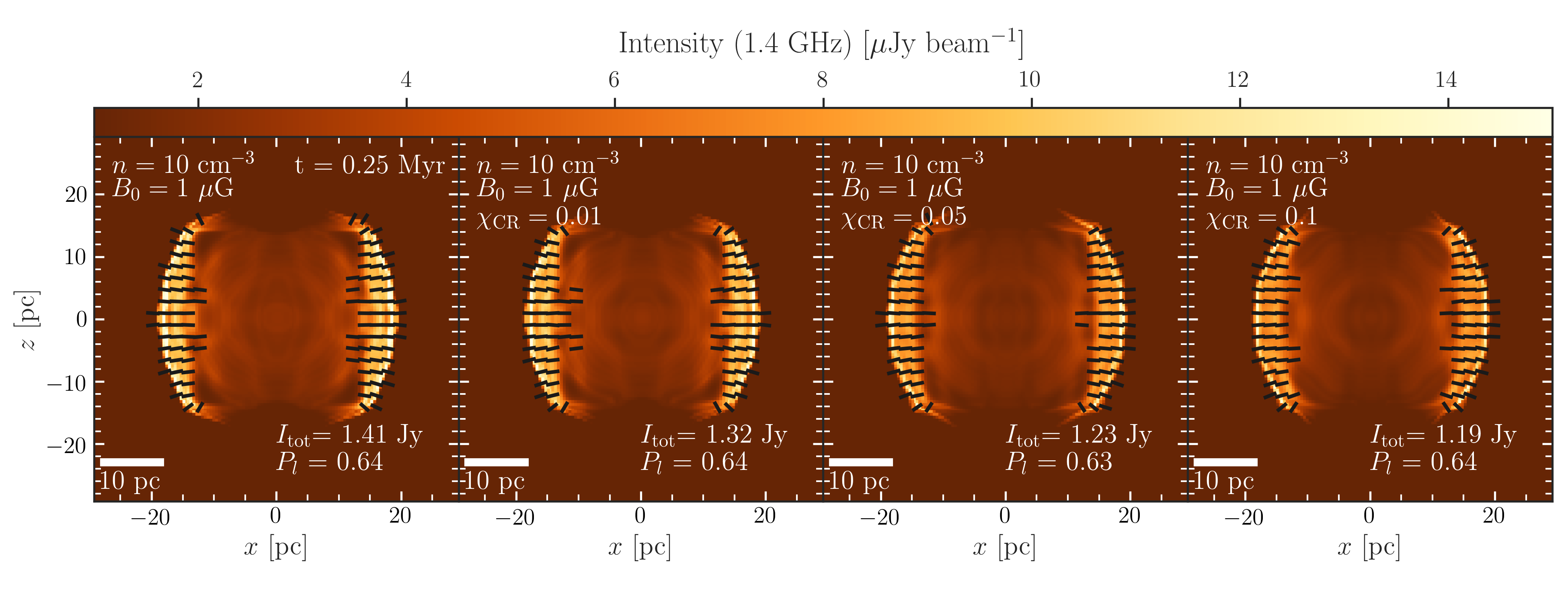}
  \caption{Synthethic synchrotron radio emission projections at 1.4~GHz for our fiducial MHD and CRMHD runs with an initially uniform magnetic field at $\sim 0.25$~Myr. Radio emission is convolved with a Gaussian of width equal to the \textit{VLA} beam-width in its B configuration. We show our runs with ambient density of $10$~cm$^{-3}$, magnetic field of 1~$\mu$G, and increasing $\chi_{\rm CR} = 0.0, 0.01, 0.05, 0.1$ from left to right, respectively.}
  \label{fig:synchro_comp_aligned}
\end{figure*}

In this section we make use of the publicly available code {\sc polaris}\footnote{\href{https://www1.astrophysik.uni-kiel.de/~polaris/}{https://www1.astrophysik.uni-kiel.de/~polaris/}} \citep{Reissl2016,Reissl2019} to generate mock radio synchrotron observations of our SNRs. To produce these maps, we provide {\sc polaris} with all the required data interpolating our simulation into a {\sc polaris} grid with double the local AMR resolution. In order to account for unresolved magnetic amplification \citep{Ji2016} and magnetic energy below the scale of the grid \citep{Schekochihin2002}, we boost the magnetic field strength by 2 dex. This yields a maximum magnetic field of $\sim$mG, comparable with observations \citep{Inoue2009}. For those quantities not included in our simulations, we follow an approach similar to \citet{Reissl2019}. We bound the electronic CRs Lorentz factor between $\gamma_\text{min} = 4$ \citep{Webber1998} and $\gamma_\text{max} \gg 1$ ($\gamma_\text{max} = 300$). Similarly, we fix the power-law index to $p = 3$ \citep{MivilleDeschenes2008}, where this power-law index is related with the spectral index $\alpha$ as $\alpha = (p - 1)/2 $.  Finally, we use the CR2 model by \citet{Reissl2019} to obtain the energy density distribution of electronic cosmic rays $e_{e^{-,{\rm CR}}}$, which assumes energy equipartition with the local magnetic energy.

{\sc polaris} generates fixed resolution maps for each Stokes vector component $\vec{S}=(I, Q, U, V)^{T}$, where $I$ is the total intensity, $Q$ and $U$ characterise the linearly polarised intensity, and $V$ the circularly polarised intensity. We present in Fig. \ref{fig:synchro_comp_aligned} the results for the fiducial CRMHD runs in the 1.4 GHz radio band, with the colourmap showing the total intensity $I$, and colour scale limits shared across all panels. The polarisation information is represented as black quivers that sample regions of $\sim2$ pc. Their inclination is given by the direction of the linearly polarised emission:
\begin{align}\label{eq:pol_angle}
    \tan{2\theta} = \frac{U}{Q},
\end{align}
and their lengths to the linear polarisation fraction:
\begin{align}\label{eq:linear_pol_fraction}
    P_l = \frac{\sqrt{U^2 + Q^2}}{I} \in [0,1].
\end{align}
The direction of these polarisation vectors is thus perpendicular to the local orientation of the magnetic field. We only show this polarisation quivers in those regions where $I$ is above 25\% of its maximum value. To better reproduce real radio interferometry observation, we convolve our emission maps with a synthesized beamwidth $\theta_{\text{HPBW}}$ Gaussian beam. We match the \textit{Very Large Array} (\textit{VLA}) radio interferometry telescope specifications in its B configuration (\textit{VLA} Observational Status Summary 2021A \citep{Momjian2017}). We assume a distance to the SNR of $\sim 3$ kpc. Note that at 0.25 Myr we are observing a significantly older SNR compared with the typical population of radio observations \citep[e.g.][]{West2015, West2017}.

Fig. \ref{fig:synchro_comp_aligned} shows a decrease of luminosity of the equatorial rims as $\chi_{\rm CR}$ increases from left to right. This decrease reaches $\sim$16\% for $\chi_{\rm CR}=0.1$. This is due to the lower magnetic field along the $x$ direction as higher CR pressure provides supports against shock compression. Thus, the CR2 model from \citet{Reissl2019} yields lower CR electron density and synchrotron luminosity. In runs with cosmic rays, the CR-driven outflows along the magnetic axis have only a minor imprint on the emission close to the polar caps. Differences are below a percent of the MHD value in these tips for the run with $\chi_{\rm CR}$=0.01, but grow to $\geq 10$\% with $\chi_{\rm CR}$=0.1. Outflows have negligible emission in this radio waveband due to low density and magnetic energy. Our fiducial MHD and CRMHD runs all reproduce the bipolar morphology observed in old SNe such as e.g. SN1006 \citep[e.g.][]{Gaensler1997,Reynoso2013,Katsuda2017}, and integrated luminosities of $\sim 1$ Jy, comparable to those obtained by \textit{VLA} studies in the 1.4-1.49 GHz range \citep[e.g.][]{Reynolds1993,Reynoso1997,Reich2002}.

The spatial resolution of our simulations is not enough to capture significant turbulent amplification. Thus compression is the dominant mechanism driving magnetic field growth behind the shock. As the CR2 model assumes equipartition between magnetic energy and electronic CRs, it follows that the emission in our mock maps is dominated by the shock and post-shock regions. The linear polarisation angle in Fig. \ref{fig:synchro_comp_aligned} tilts along the shell and remains radially aligned. This indicates a tangential magnetic field wrapping the shell, hence reproducing the radial polarisation seen in previous work \citep{Fulbright1990,West2017}. The polarisation fraction $P_l$ appears to remain practically constant when including CRs, although approaching the theoretical maximum of 70\% rarely seen in observations \citep[e.g.][]{Gaensler1997,DeLaney2002}.

\begin{figure*}
  \centering \includegraphics[width=\textwidth]{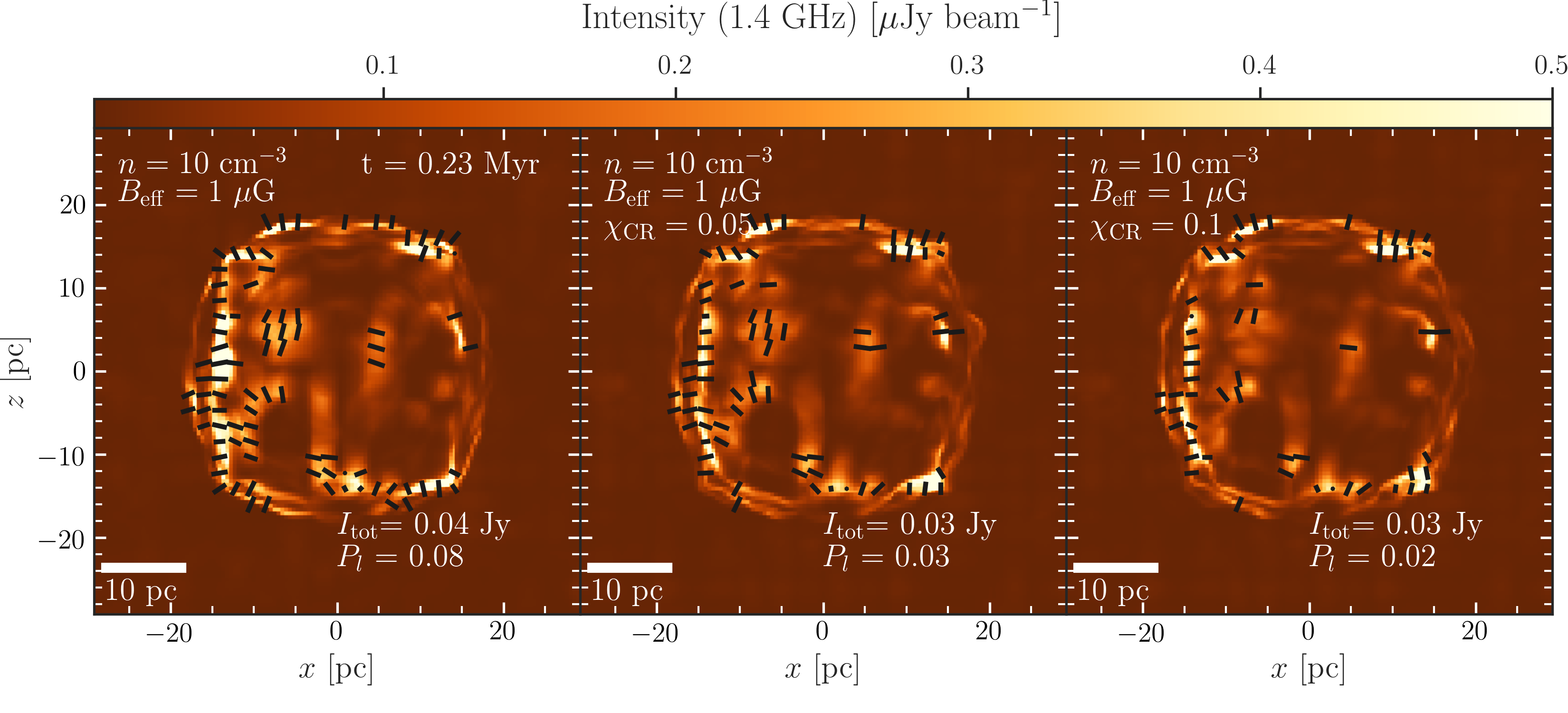}
  \caption{Projected synthethic synchrotron radio emission in the 1.4 GHz band for our fiducial MHD and CRMHD runs with a tangled initial configuration of the magnetic field at $\sim 0.25$ Myr. Maps are convolved with a Gaussian of width equivalent to the \textit{VLA} beamwidth in its B configuration. Panels show our runs with 10 cm$^{-3}$, $B_\text{eff} = 1\mu$G, and with $\chi_{\rm CR}= 0.0$ (left), 0.05 (centre) and 0.1 (right).}
  \label{fig:synchro_comp_turb}
\end{figure*}

We also review mock synchrotron observations for our simulations employing tangled initial conditions for the magnetic fields in Fig.~\ref{fig:synchro_comp_turb}. The SNR shell is the main contributor to the synchrotron emission, as observed when compared with the density slices of Fig.~\ref{fig:slices_magics} (leftmost column). These projections also display a decrease in luminosity with increasing $\chi_{\rm CR}$, making this result independent of the ambient magnetic field configuration. The synthetic observations present filaments and patches of emission emerging from the post-shock region, evidencing that the tangled magnetic field initial conditions imprint signatures on the synchrotron emission morphology and its luminosity. Besides these peculiarities, the shell emission remains approximately radially polarised, albeit to a lesser extent than for the uniform case. Thus, the magnetic field remains as before, wrapped approximately tangentially around the SNR shock, matching observations \citep[e.g.][]{Milne1987, Anderson1995,Reynoso1997}. This agrees with the conclusions of \cite{West2017}, in which a random magnetic field configuration and quasi-parallel CR injection still produce radial polarisation along the shell of SNR due to the CR distribution tracing the projected magnetic field. Opposite to what was found for the aligned runs, these maps present $P_l$ of the order of a few percents. Therefore, these maps argue in favour of a significant fraction of the magnetic energy in SNRs residing in a turbulent component \citep{Reynolds1993,DeLaney2002,West2017}.

\section{Discussion and Caveats}\label{sec:discussion}
We have reviewed the late time evolution of SNRs when accounting for cosmic ray magneto-hydrodynamics. We have explored various values of ambient number density, initial magnetic field strengths and configuration, as well as initial fraction of SN energy in CRs. While we account for additional physics frequently not included in standard SNR simulations, our models still have some limitations. We discuss these and the main caveats of our work here.

Our tangled magnetic field initial conditions provide a more realistic setup for the propagation of CRs in and around an evolving SNR than the uniform magnetic field setup. However, they are evolved in a uniform density and static environment, in which magnetic inhomogeneities will drive turbulent perturbations with time. The evolution of individual SNR has been studied in inhomogenoeus media, e.g. adopting two-phase initial conditions \citep[][]{Kim2015} or supersonic turbulence \citep[][]{Iffrig2014, Martizzi2015}. The momentum deposition of SNe was found to depend weakly on environmental properties, consistent with previous work in homogeneous ambient media \citep{Chevalier1974,Cioffi1988,Blondin1998}. However, these studies only considered simulations of SNRs without CRs. Our CRMHD runs find that a significant contribution to the momentum deposited by the SN comes from escaping CRs travelling through the ambient medium. CRs suffer lower hadronic and Coulomb cooling energy losses in a low density medium. Similarly, our uniform CRMHD simulations have shown that their boost to the momentum is more significant at lower densities. Equally, the impact of escaping CRs will depend on ambient instabilities and structure. Consequently, modelling of a realistic ISM may be important for the deposition of momentum by CRs in the ambient medium while being secondary for the SNR momentum. Furthermore, our parameter-space is limited to one metallicity. As cooling functions depend on gas metallicity, this quantity influences the evolution of the SNR. High metallicities increase gas radiative cooling and hence reduce the final SNR momentum \citep[e.g.][]{Martizzi2015}. Regardless, we expect the overall behaviour observed with metallicity to remain unchanged by CRs.

Numerical simulations have evolved isolated individual SNRs to study their effects at late times in their surrounding ISM. Nevertheless, the efficiency of SNe feedback depends on the pre-processing of the ambient medium by radiation \citep{Geen2014, Walch2015}, stellar winds \citep{Geen2014}, and previous SNe \citep{Hennebelle2014, Walch2015}. These processes interact with each other in a non-trivial manner, affecting the conditions of the ambient medium in which SNRs evolve. Additionally, while classical SNRs are limited to influence the region reached by their forward shock, our CRMHD simulations show that CRs originated in the SN influence the ambient gas ahead and far from the shock, accelerating gas and smoothing out ambient inhomegeneities in the process. We show that CRs are able to drive further expansion of SNRs at late times, in agreement with theoretical predictions \citep{Diesing2018}. Our work suggests that CRs extend the influence of SNRs beyond their immediate vicinity, which will play a role in the evolution of neighbouring SNe as well as affect larger galactic scales \citep{Hanasz2013,Salem2014,Chan2019,Dashyan2020}.

We initially inject CRs altogether with thermal energy in the SNe at the centre of the grid. There are two main reasons justifying this method. Firstly, that the scales in which CR acceleration takes place are below our grid resolution (e.g. for a proton with an energy of $\sim 1$ GeV in a magnetic field of 1 $\mu$G the gyroradius is $\sim 10^{-6}$ pc). Secondly, the shock is less efficient in the acceleration of high energy CRs at the times studied here \citep[e.g.][]{Ferriere1991b, Caprioli2014}. Furthermore, hydrodynamical simulations of young SNRs with CRs \citep[e.g.][]{Pfrommer2017, Pais2018a, Dubois2019} have shown that the self-similarity of Sedov-Taylor solution also applies in this early evolution when considering a gas formed by a thermal-CR mixture. This is consistent with our CRMHD simulation results. However, the late time effects of CR acceleration in shocks have not been explored in previous numerical simulations. \citet{Diesing2018} have found in their semi-analytical work that the late time CR acceleration at the shock drives further expansion of the SNR, reaching an increased momentum deposition of a factor of $\sim 2-3$, much higher than the ones found in this work. CR acceleration can affect the characteristics of the shock \citep[e.g.][]{Ferriere1991b,Girichidis2014, Pfrommer2017,Dubois2019}, which in turn could affect the acceleration of the SNR at late times. This suggests the need for the exploration of CR acceleration well after the Sedov-Taylor phase self-consistently injecting CRs by employing a shock finder similar to those used by \citet{Pfrommer2017,Dubois2019}. Also, we restrict ourselves to the low end of diffusion values probed by galaxy formation simulations \citep[e.g.][]{Uhlig2012,Booth2013,Salem2014,Girichidis2016,Chan2019,Dashyan2020}. However, the complexity of CR transport models used in numerical experiments remains unconstrained by existing observations, which are still consistent with a simple isotropic diffusion model with $D_0\sim 10^{28}$ cm$^2$ s$^{-1}$ \citep[e.g.][and references therein]{Evoli2019}. Additionally, the cosmic ray diffusion rate is unlikely to be constant. For example, regions of high star-formation permeated with significant turbulence driven by stellar feedback have stronger and turbulent magnetic fields at scales below those resolved by galaxy formation simulations. Such magnetic turbulence would lead to an enhanced confinement of CRs, and thus a local reduction of $D$ several orders of magnitude below the Galactic average \citep{Schroer2020,Semenov2021}. Our value of $D_0= 3\times 10^{27}\, \rm cm^2\,s^{-1}$ is hence a compromise between this small-scales suppressed value and the galactic average. Nonetheless, it remains interesting to explore lower/higher diffusion coefficients and how they may affect the momentum deposition of SNe with CRs.

\section{Conclusions}\label{sec:conclusions}
In this work, we study HD, MHD and CRMHD simulations of individual SNe expanding in a uniform medium in order to explore the influence of CRs in their evolution and momentum deposition. Our suite explores the parameter space across three initial configuration variables: gas number density (from 1 to 100 cm$^{-3}$), magnetic field strength (0.1, 1 and 5 $\mu$G; exploring both uniform and tangled magnetic field configurations) and fraction of SN energy injected as CRs energy (1\%, 5\% and 10\%). Our simulations capture the evolution from the start of the Sedov-Taylor phase until the convergence of momentum deposition. We also review the morphological appearance of our SNR in synchrotron emission maps generated using the publicly available code {\sc polaris} \citep{Reissl2016,Reissl2019}. Our main findings are as follow:
\begin{itemize}
    \item In agreement with previous work, we find no significant difference in the final outward radial momentum of SNe when including an initially aligned magnetic field with respect to the pure HD simulations, regardless of its strength within our explored range. Nonetheless, increasing $\vert \vec{B}\vert$ stabilises and widens the SNR shell, reducing the volume of hot gas.
    \item Cosmic rays affect the morphology of our SNRs depending on the initial configuration of the magnetic field. When employing a uniform magnetic field, we find cosmic rays to escape along the poles of the remnants and accelerate the gas beyond the shell. SNRs in our simulations with a tangled initial magnetic field resemble the morphology of the HD case, with cosmic rays escaping beyond the shock isotropically.
    \item Once radiative losses become significant, runs with CRs present a different evolution. The energy dominance in the bubble shifts from thermal to CRs, and we find an additional \textit{CR pressure-driven snowplough} phase, characterised by a further gain in momentum of up to $\sim 50$\% in the run with $\chi_{\text{CR}}=0.1$ by 4 Myr. The total outward momentum when using our tangled magnetic field initial conditions is even slightly higher than for the CRMHD runs with an aligned magnetic field.
    \item We therefore revise the momentum deposition law, accounting for the `boost' due to CR injection: $p = 2.87\times 10^{5} (\chi_{\text{CR}} + 1)^{4.82}\left(\frac{n}{\text{cm}^{-3}}\right)^{-0.196} M_{\odot}$ km s$^{-1}$. We find an inverse dependence with the number density, in contrast with previous analytical studies \citep{Diesing2018}.
    \item We find that an initially aligned magnetic field is able to reproduce the bipolar, radially polarised emission structure of old SNRs. The dynamical effects of CR pressure in the equatorial shock and in the polar `outflows' imprint reductions in the synchrotron luminosity directly proportional to $\chi_{\rm CR}$. Our runs with a tangled magnetic field configuration reproduce instead the radial polarisation of young SNRs, consistent with the results of \cite{West2017} and with a lower linear polarisation fraction than the aligned case.
\end{itemize}

Overall, we find CRs to be a key factor in the late time evolution of SNRs. Their effects extend well beyond the SNR shock into the SNR surroundings and wider interstellar medium. These results indicate that in future work it will be important to explore the additional deposition of momentum due local and escaping CRs at galactic scales which may affect significantly the thermodynamics of galaxy-scale outflows.   

\section*{Acknowledgements}

FRM is supported by the Wolfson Harrison UK Research Council Physics Scholarship. This work was supported by the ERC Starting Grant 638707 `Black holes and their host galaxies: co-evolution across cosmic time" and by STFC. This work was performed using resources provided by the Cambridge Service for Data Driven Discovery (CSD3) operated by the University of Cambridge Research Computing Service, provided by Dell EMC and Intel using Tier-2 funding from the Engineering and Physical Sciences Research Council (capital grant EP/P020259/1), and DiRAC funding from the Science and Technology Facilities Council. Visualisations and analysis codes in this paper were based on the {\sc yt project} \citep{Turk2011}.

\section*{Data Availability}

The simulation data underlying this article will be shared on reasonable request to the corresponding author.



\bibliographystyle{mnras}
\bibliography{references} 

\begin{thebibliography}{}
\makeatletter
\relax
\def\mn@urlcharsother{\let\do\@makeother \do\$\do\&\do\#\do\^\do\_\do\%\do\~}
\def\mn@doi{\begingroup\mn@urlcharsother \@ifnextchar [ {\mn@doi@}
  {\mn@doi@[]}}
\def\mn@doi@[#1]#2{\def\@tempa{#1}\ifx\@tempa\@empty \href
  {http://dx.doi.org/#2} {doi:#2}\else \href {http://dx.doi.org/#2} {#1}\fi
  \endgroup}
\def\mn@eprint#1#2{\mn@eprint@#1:#2::\@nil}
\def\mn@eprint@arXiv#1{\href {http://arxiv.org/abs/#1} {{\tt arXiv:#1}}}
\def\mn@eprint@dblp#1{\href {http://dblp.uni-trier.de/rec/bibtex/#1.xml}
  {dblp:#1}}
\def\mn@eprint@#1:#2:#3:#4\@nil{\def\@tempa {#1}\def\@tempb {#2}\def\@tempc
  {#3}\ifx \@tempc \@empty \let \@tempc \@tempb \let \@tempb \@tempa \fi \ifx
  \@tempb \@empty \def\@tempb {arXiv}\fi \@ifundefined
  {mn@eprint@\@tempb}{\@tempb:\@tempc}{\expandafter \expandafter \csname
  mn@eprint@\@tempb\endcsname \expandafter{\@tempc}}}

\bibitem[\protect\citeauthoryear{Acciari et~al.,}{Acciari
  et~al.}{2009}]{Acciari2009}
Acciari V.~A.,  et~al., 2009, \mn@doi [Nature] {10.1038/nature08557}, 462, 770

\bibitem[\protect\citeauthoryear{Anderson, Keohane  \& Rudnick}{Anderson
  et~al.}{1995}]{Anderson1995}
Anderson M.~C.,  Keohane J.~W.,   Rudnick L.,  1995, \mn@doi [ApJ]
  {10.1086/175356}, 441, 300

\bibitem[\protect\citeauthoryear{Aumer, White, Naab  \& Scannapieco}{Aumer
  et~al.}{2013}]{Aumer2013}
Aumer M.,  White S. D.~M.,  Naab T.,   Scannapieco C.,  2013, \mn@doi [MNRAS]
  {10.1093/mnras/stt1230}, 434, 3142

\bibitem[\protect\citeauthoryear{Beck}{Beck}{2001}]{Beck2001}
Beck R.,  2001, \mn@doi [Space Sci. Rev.] {10.1023/A:1013805401252}, 99, 243

\bibitem[\protect\citeauthoryear{Beck}{Beck}{2007}]{Beck2007}
Beck R.,  2007, \mn@doi [A{\&}A] {10.1051/0004-6361:20066988}, 470, 539

\bibitem[\protect\citeauthoryear{Beck, {Beck}  \& {Rainer}}{Beck
  et~al.}{2015}]{Beck2015}
Beck R.,  {Beck}  {Rainer} 2015, \mn@doi [A{\&}ARv]
  {10.1007/S00159-015-0084-4}, 24, 4

\bibitem[\protect\citeauthoryear{Behroozi, Wechsler  \& Conroy}{Behroozi
  et~al.}{2013}]{Behroozi2013}
Behroozi P.~S.,  Wechsler R.~H.,   Conroy C.,  2013, \mn@doi [ApJ]
  {10.1088/0004-637X/770/1/57}, 770, 57

\bibitem[\protect\citeauthoryear{Bhat \& Subramanian}{Bhat \&
  Subramanian}{2013}]{Bhat2013}
Bhat P.,  Subramanian K.,  2013, \mn@doi [MNRAS] {10.1093/mnras/sts516}, 429,
  2469

\bibitem[\protect\citeauthoryear{Blondin, Wright, Borkowski  \&
  Reynolds}{Blondin et~al.}{1998}]{Blondin1998}
Blondin J.~M.,  Wright E.~B.,  Borkowski K.~J.,   Reynolds S.~P.,  1998,
  \mn@doi [ApJ] {10.1086/305708}, 500, 342

\bibitem[\protect\citeauthoryear{Booth, Schaye, Delgado  \&
  Dalla~Vecchia}{Booth et~al.}{2012}]{Booth2012}
Booth C.~M.,  Schaye J.,  Delgado J.~D.,   Dalla~Vecchia C.,  2012, \mn@doi
  [MNRAS] {10.1111/j.1365-2966.2011.20047.x}, 420, 1053

\bibitem[\protect\citeauthoryear{Booth, Agertz, Kravtsov  \& Gnedin}{Booth
  et~al.}{2013}]{Booth2013}
Booth C.~M.,  Agertz O.,  Kravtsov A.~V.,   Gnedin N.~Y.,  2013, \mn@doi [ApJ]
  {10.1088/2041-8205/777/1/L16}, 777, L16

\bibitem[\protect\citeauthoryear{Brahimi, Marcowith"  \& Ptuskin"}{Brahimi
  et~al.}{2019}]{Brahimi2019}
Brahimi L.,  Marcowith" A.,   Ptuskin" V.,  2019, in Proceedings of 36th
  International Cosmic Ray Conference — PoS(ICRC2019). Sissa Medialab,
  Trieste, Italy, p.~040, \mn@doi{10.22323/1.358.0040}

\bibitem[\protect\citeauthoryear{Caprioli \& Spitkovsky}{Caprioli \&
  Spitkovsky}{2014}]{Caprioli2014}
Caprioli D.,  Spitkovsky A.,  2014, \mn@doi [ApJ] {10.1088/0004-637X/794/1/47},
  794, 47

\bibitem[\protect\citeauthoryear{Chan, Kere{\v{s}}, Hopkins, Quataert, Su,
  Hayward  \& Faucher-Gigu{\`{e}}re}{Chan et~al.}{2019}]{Chan2019}
Chan T.~K.,  Kere{\v{s}} D.,  Hopkins P.~F.,  Quataert E.,  Su K.-Y.,  Hayward
  C.~C.,   Faucher-Gigu{\`{e}}re C.-A.,  2019, \mn@doi [MNRAS]
  {10.1093/mnras/stz1895}, 488, 3716

\bibitem[\protect\citeauthoryear{Chevalier}{Chevalier}{1974}]{Chevalier1974}
Chevalier R.~A.,  1974, \mn@doi [ApJ] {10.1086/152740}, 188, 501

\bibitem[\protect\citeauthoryear{Chevalier}{Chevalier}{1983}]{Chevalier1983}
Chevalier R.~A.,  1983, \mn@doi [ApJ] {10.1086/161338}, 272, 765

\bibitem[\protect\citeauthoryear{Cioffi, McKee  \& Bertschinger}{Cioffi
  et~al.}{1988}]{Cioffi1988}
Cioffi D.~F.,  McKee C.~F.,   Bertschinger E.,  1988, \mn@doi [ApJ]
  {10.1086/166834}, 334, 252

\bibitem[\protect\citeauthoryear{Crutcher, Wandelt, Heiles, Falgarone  \&
  Troland}{Crutcher et~al.}{2010}]{Crutcher2010}
Crutcher R.~M.,  Wandelt B.,  Heiles C.,  Falgarone E.,   Troland T.~H.,  2010,
  \mn@doi [ApJ] {10.1088/0004-637X/725/1/466}, 725, 466

\bibitem[\protect\citeauthoryear{Dalla~Vecchia \& Schaye}{Dalla~Vecchia \&
  Schaye}{2012}]{DallaVecchia2012}
Dalla~Vecchia C.,  Schaye J.,  2012, \mn@doi [MNRAS]
  {10.1111/j.1365-2966.2012.21704.x}, 426, 140

\bibitem[\protect\citeauthoryear{Dashyan \& Dubois}{Dashyan \&
  Dubois}{2020}]{Dashyan2020}
Dashyan G.,  Dubois Y.,  2020, \mn@doi [A{\&}A] {10.1051/0004-6361/201936339},
  638, 123

\bibitem[\protect\citeauthoryear{DeLaney, Koralesky, Rudnick  \&
  Dickel}{DeLaney et~al.}{2002}]{DeLaney2002}
DeLaney T.,  Koralesky B.,  Rudnick L.,   Dickel J.~R.,  2002, \mn@doi [ApJ]
  {10.1086/343787}, 580, 914

\bibitem[\protect\citeauthoryear{Dermer \& Powale}{Dermer \&
  Powale}{2013}]{Dermer2013}
Dermer C.~D.,  Powale G.,  2013, \mn@doi [A{\&}A]
  {10.1051/0004-6361/201220394}, 553, 34

\bibitem[\protect\citeauthoryear{Diesing \& Caprioli}{Diesing \&
  Caprioli}{2018}]{Diesing2018}
Diesing R.,  Caprioli D.,  2018, \mn@doi [Phys. Rev. Lett.]
  {10.1103/PhysRevLett.121.091101}, 121, 91101

\bibitem[\protect\citeauthoryear{Dijkstra \& Loeb}{Dijkstra \&
  Loeb}{2008}]{Dijkstra2008}
Dijkstra M.,  Loeb A.,  2008, \mn@doi [MNRAS]
  {10.1111/j.1365-2966.2008.13920.x}, 391, 457

\bibitem[\protect\citeauthoryear{Dobbs, Burkert  \& Pringle}{Dobbs
  et~al.}{2011}]{Dobbs2011}
Dobbs C.~L.,  Burkert A.,   Pringle J.~E.,  2011, \mn@doi [MNRAS]
  {10.1111/j.1365-2966.2011.19346.x}, 417, 1318

\bibitem[\protect\citeauthoryear{Dubois \& Commer{\c{c}}on}{Dubois \&
  Commer{\c{c}}on}{2016}]{Dubois2016}
Dubois Y.,  Commer{\c{c}}on B.,  2016, \mn@doi [A{\&}A]
  {10.1051/0004-6361/201527126}, 585, 1

\bibitem[\protect\citeauthoryear{Dubois, Commer{\c{c}}on, Marcowith  \&
  Brahimi}{Dubois et~al.}{2019}]{Dubois2019}
Dubois Y.,  Commer{\c{c}}on B.,  Marcowith A.,   Brahimi L.,  2019, \mn@doi
  [A{\&}A] {10.1051/0004-6361/201936275}, 631, A121

\bibitem[\protect\citeauthoryear{Ehlert et~al.,}{Ehlert
  et~al.}{2018}]{Ehlert2018}
Ehlert K.,  et~al., 2018, \mn@doi [MNRAS] {10.1093/MNRAS/STY2397}, 481, 2878

\bibitem[\protect\citeauthoryear{Emerick, Bryan  \& Mac~Low}{Emerick
  et~al.}{2018}]{Emerick2018}
Emerick A.,  Bryan G.~L.,   Mac~Low M.-M.,  2018, \mn@doi [ApJ]
  {10.3847/2041-8213/aae315}, 865, L22

\bibitem[\protect\citeauthoryear{Everett, Zweibel, Benjamin, McCammon, Rocks
  \& Gallagher~III}{Everett et~al.}{2008}]{Everett2008}
Everett J.~E.,  Zweibel E.~G.,  Benjamin R.~A.,  McCammon D.,  Rocks L.,
  Gallagher~III J.~S.,  2008, \mn@doi [ApJ] {10.1086/524766}, 674, 258

\bibitem[\protect\citeauthoryear{Evoli, Aloisio  \& Blasi}{Evoli
  et~al.}{2019}]{Evoli2019}
Evoli C.,  Aloisio R.,   Blasi P.,  2019, \mn@doi [Phys. Rev. D]
  {10.1103/PhysRevD.99.103023}, 99, 103023

\bibitem[\protect\citeauthoryear{Federrath, Schober, Bovino  \&
  Schleicher}{Federrath et~al.}{2014}]{Federrath2014}
Federrath C.,  Schober J.,  Bovino S.,   Schleicher D.~R.,  2014, \mn@doi [ApJ]
  {10.1088/2041-8205/797/2/L19}, 797, L19

\bibitem[\protect\citeauthoryear{Ferland, Korista, Verner, Ferguson, Kingdon
  \& Verner}{Ferland et~al.}{1998}]{Ferland1998}
Ferland G.~J.,  Korista K.~T.,  Verner D.~A.,  Ferguson J.~W.,  Kingdon J.~B.,
   Verner E.~M.,  1998, \mn@doi [Publ. Astron. Soc. Pac.] {10.1086/316190},
  110, 761

\bibitem[\protect\citeauthoryear{Ferri{\`{e}}re}{Ferri{\`{e}}re}{2001}]{Ferriere2001}
Ferri{\`{e}}re K.~M.,  2001, \mn@doi [Rev. Mod. Phys.]
  {10.1103/RevModPhys.73.1031}, 73, 1031

\bibitem[\protect\citeauthoryear{Ferriere \& Zweibel}{Ferriere \&
  Zweibel}{1991}]{Ferriere1991b}
Ferriere K.~M.,  Zweibel E.~G.,  1991, \mn@doi [ApJ] {10.1086/170818}, 383, 602

\bibitem[\protect\citeauthoryear{Ferriere, Mac~Low  \& Zweibel}{Ferriere
  et~al.}{1991}]{Ferriere1991a}
Ferriere K.~M.,  Mac~Low M.-M.,   Zweibel E.~G.,  1991, \mn@doi [ApJ]
  {10.1086/170185}, 375, 239

\bibitem[\protect\citeauthoryear{Fromang, Hennebelle  \& Teyssier}{Fromang
  et~al.}{2006}]{Fromang2006}
Fromang S.,  Hennebelle P.,   Teyssier R.,  2006, \mn@doi [A{\&}A]
  {10.1051/0004-6361:20065371}, 457, 371

\bibitem[\protect\citeauthoryear{Fulbright \& Reynolds}{Fulbright \&
  Reynolds}{1990}]{Fulbright1990}
Fulbright M.~S.,  Reynolds S.~P.,  1990, \mn@doi [ApJ] {10.1086/168947}, 357,
  591

\bibitem[\protect\citeauthoryear{Gaensler}{Gaensler}{1997}]{Gaensler1997}
Gaensler B.~M.,  1997, \mn@doi [ApJ] {10.1086/305146}, 493, 781

\bibitem[\protect\citeauthoryear{Geen, Rosdahl, Blaizot, Devriendt  \&
  Slyz}{Geen et~al.}{2014}]{Geen2014}
Geen S.,  Rosdahl J.,  Blaizot J.,  Devriendt J.,   Slyz A.,  2014, \mn@doi
  [MNRAS] {10.1093/mnras/stv251}, 448, 3248

\bibitem[\protect\citeauthoryear{Geen, Hennebelle, Tremblin  \& Rosdahl}{Geen
  et~al.}{2016}]{Geen2016}
Geen S.,  Hennebelle P.,  Tremblin P.,   Rosdahl J.,  2016, \mn@doi [MNRAS]
  {10.1093/mnras/stw2235}, 463, 3129

\bibitem[\protect\citeauthoryear{Girichidis, Naab, Walch  \& Hanasz}{Girichidis
  et~al.}{2014}]{Girichidis2014}
Girichidis P.,  Naab T.,  Walch S.,   Hanasz M.,  2014, arXiv:1406.4861
  [astro-ph.HE]

\bibitem[\protect\citeauthoryear{Girichidis et~al.,}{Girichidis
  et~al.}{2016}]{Girichidis2016}
Girichidis P.,  et~al., 2016, \mn@doi [ApJ] {10.3847/2041-8205/816/2/l19}, 816,
  L19

\bibitem[\protect\citeauthoryear{Girichidis, Naab, Hanasz  \& Walch}{Girichidis
  et~al.}{2018}]{Girichidis2018}
Girichidis P.,  Naab T.,  Hanasz M.,   Walch S.,  2018, \mn@doi [MNRAS]
  {10.1093/mnras/sty1653}, 479, 3042

\bibitem[\protect\citeauthoryear{Hanasz, Lesch, Naab, Gawryszczak, Kowalik  \&
  W{\'{o}}lta{\'{n}}ski}{Hanasz et~al.}{2013}]{Hanasz2013}
Hanasz M.,  Lesch H.,  Naab T.,  Gawryszczak A.,  Kowalik K.,
  W{\'{o}}lta{\'{n}}ski D.,  2013, \mn@doi [ApJ] {10.1088/2041-8205/777/2/L38},
  777, L38

\bibitem[\protect\citeauthoryear{Hanayama \& Tomisaka}{Hanayama \&
  Tomisaka}{2006}]{Hanayama2006}
Hanayama H.,  Tomisaka K.,  2006, \mn@doi [ApJ] {10.1086/500527}, 641, 905

\bibitem[\protect\citeauthoryear{Helder et~al.,}{Helder
  et~al.}{2013}]{Helder2013}
Helder E.~A.,  et~al., 2013, \mn@doi [ApJ] {10.1088/0004-637X/764/1/11}, 764,
  11

\bibitem[\protect\citeauthoryear{Hennebelle \& Iffrig}{Hennebelle \&
  Iffrig}{2014}]{Hennebelle2014}
Hennebelle P.,  Iffrig O.,  2014, \mn@doi [A{\&}A]
  {10.1051/0004-6361/201423392}, 570

\bibitem[\protect\citeauthoryear{Hopkins, Quataert  \& Murray}{Hopkins
  et~al.}{2011}]{Hopkins2011}
Hopkins P.~F.,  Quataert E.,   Murray N.,  2011, \mn@doi [MNRAS]
  {10.1111/j.1365-2966.2011.19306.x}, 417, 950

\bibitem[\protect\citeauthoryear{Hopkins et~al.,}{Hopkins
  et~al.}{2018}]{Hopkins2018}
Hopkins P.~F.,  et~al., 2018, \mn@doi [MNRAS] {10.1093/mnras/sty1690}, 480, 800

\bibitem[\protect\citeauthoryear{Hopkins, Chan, Ji, Hummels, Kereˇs, Quataert
  \& Faucher-Gigu{\`{e}}re}{Hopkins et~al.}{2021}]{Hopkins2021}
Hopkins P.~F.,  Chan T.~K.,  Ji S.,  Hummels C.~B.,  Kereˇs D.,  Quataert E.,
   Faucher-Gigu{\`{e}}re C.~A.,  2021, \mn@doi [MNRAS]
  {10.1093/mnras/staa3690}, 501, 3640

\bibitem[\protect\citeauthoryear{Iffrig \& Hennebelle}{Iffrig \&
  Hennebelle}{2014}]{Iffrig2014}
Iffrig O.,  Hennebelle P.,  2014, \mn@doi [A{\&}A]
  {10.1051/0004-6361/201424556}, 576, 1

\bibitem[\protect\citeauthoryear{Inoue, Yamazaki  \& Inutsuka}{Inoue
  et~al.}{2009}]{Inoue2009}
Inoue T.,  Yamazaki R.,   Inutsuka S.~I.,  2009, \mn@doi [ApJ]
  {10.1088/0004-637X/695/2/825}, 695, 825

\bibitem[\protect\citeauthoryear{Jansson \& Farrar}{Jansson \&
  Farrar}{2012}]{Jansson2012}
Jansson R.,  Farrar G.~R.,  2012, \mn@doi [ApJ] {10.1088/0004-637X/757/1/14},
  757, 14

\bibitem[\protect\citeauthoryear{Ji, Oh, Ruszkowski, Markevitch, Peng~Oh,
  Ruszkowski  \& Markevitch}{Ji et~al.}{2016}]{Ji2016}
Ji S.,  Oh S.~P.,  Ruszkowski M.,  Markevitch M.,  Peng~Oh S.,  Ruszkowski M.,
   Markevitch M.,  2016, \mn@doi [MNRAS] {10.1093/mnras/stw2320}, 463, 3989

\bibitem[\protect\citeauthoryear{Jubelgas, Springel, En{\ss}lin  \&
  Pfrommer}{Jubelgas et~al.}{2008}]{Jubelgas2008}
Jubelgas M.,  Springel V.,  En{\ss}lin T.,   Pfrommer C.,  2008, \mn@doi
  [A{\&}A] {10.1051/0004-6361:20065295}, 481, 33

\bibitem[\protect\citeauthoryear{Katsuda}{Katsuda}{2017}]{Katsuda2017}
Katsuda S.,  2017, in , Handbook of Supernovae.
Springer International Publishing, Cham, pp 63--81,
  \mn@doi{10.1007/978-3-319-21846-5{\_}45}, \url
  {http://link.springer.com/10.1007/978-3-319-21846-5_45}

\bibitem[\protect\citeauthoryear{Katz et~al.,}{Katz et~al.}{2021}]{Katz2021}
Katz H.,  et~al., 2021, \mn@doi [MNRAS] {10.1093/mnras/stab2148}, 507, 1254

\bibitem[\protect\citeauthoryear{Kay, Pearce, Frenk  \& Jenkins}{Kay
  et~al.}{2002}]{Kay2002}
Kay S.~T.,  Pearce F.~R.,  Frenk C.~S.,   Jenkins A.,  2002, \mn@doi [MNRAS]
  {10.1046/j.1365-8711.2002.05070.x}, 330, 113

\bibitem[\protect\citeauthoryear{Kazantsev}{Kazantsev}{1968}]{Kazantsev1968}
Kazantsev A.,  1968, J. Exp. Theor. Phys., 26, 1031

\bibitem[\protect\citeauthoryear{Kim \& Ostriker}{Kim \&
  Ostriker}{2015}]{Kim2015}
Kim C.~G.,  Ostriker E.~C.,  2015, \mn@doi [ApJ] {10.1088/0004-637X/802/2/99},
  802, 99

\bibitem[\protect\citeauthoryear{Kimm, Cen, Devriendt, Dubois  \& Slyz}{Kimm
  et~al.}{2015}]{Kimm2015}
Kimm T.,  Cen R.,  Devriendt J.,  Dubois Y.,   Slyz A.,  2015, \mn@doi [MNRAS]
  {10.1093/mnras/stv1211}, 451, 2900

\bibitem[\protect\citeauthoryear{Kimm, Haehnelt, Blaizot, Katz, Michel-Dansac,
  Garel, Rosdahl  \& Teyssier}{Kimm et~al.}{2018}]{Kimm2018}
Kimm T.,  Haehnelt M.,  Blaizot J.,  Katz H.,  Michel-Dansac L.,  Garel T.,
  Rosdahl J.,   Teyssier R.,  2018, \mn@doi [MNRAS] {10.1093/mnras/sty126},
  475, 4617

\bibitem[\protect\citeauthoryear{Kulsrud \& Pearce}{Kulsrud \&
  Pearce}{1969}]{Kulsrud1969}
Kulsrud R.,  Pearce W.~P.,  1969, \mn@doi [ApJ] {10.1086/149981}, 156, 445

\bibitem[\protect\citeauthoryear{Li, Ostriker, Cen, Bryan  \& Naab}{Li
  et~al.}{2015}]{Li2015}
Li M.,  Ostriker J.~P.,  Cen R.,  Bryan G.~L.,   Naab T.,  2015, \mn@doi [ApJ]
  {10.1088/0004-637X/814/1/4}, 814, 4

\bibitem[\protect\citeauthoryear{Liu \& Gao}{Liu \& Gao}{2010}]{Liu2010}
Liu F.,  Gao Y.,  2010, \mn@doi [ApJ] {10.1088/0004-637X/713/1/524}, 713, 524

\bibitem[\protect\citeauthoryear{Martin-Alvarez, Devriendt, Slyz  \&
  Teyssier}{Martin-Alvarez et~al.}{2018}]{Martin-Alvarez2018}
Martin-Alvarez S.,  Devriendt J.,  Slyz A.,   Teyssier R.,  2018, \mn@doi
  [MNRAS] {10.1093/mnras/sty1623}, 479, 3343

\bibitem[\protect\citeauthoryear{Martin-Alvarez, Slyz, Devriendt  \&
  G{\'{o}}mez-Guijarro}{Martin-Alvarez et~al.}{2020}]{Martin-Alvarez2020}
Martin-Alvarez S.,  Slyz A.,  Devriendt J.,   G{\'{o}}mez-Guijarro C.,  2020,
  \mn@doi [MNRAS] {10.1093/mnras/staa1438}, 495, 4475

\bibitem[\protect\citeauthoryear{Martizzi, Faucher-gigu{\`{e}}re  \&
  Quataert}{Martizzi et~al.}{2015}]{Martizzi2015}
Martizzi D.,  Faucher-gigu{\`{e}}re C.~A.,   Quataert E.,  2015, \mn@doi
  [MNRAS] {10.1093/mnras/stv562}, 450, 504

\bibitem[\protect\citeauthoryear{Meaburn, Hartquist  \& Dyson}{Meaburn
  et~al.}{1988}]{Meaburn1988}
Meaburn J.,  Hartquist T.~W.,   Dyson J.~E.,  1988, \mn@doi [MNRAS]
  {10.1093/mnras/230.2.243}, 230, 243

\bibitem[\protect\citeauthoryear{Milne, Caswell, Haynes, Kesteven, Wellington
  \& Roger}{Milne et~al.}{1987}]{Milne1987}
Milne D.,  Caswell J.,  Haynes R.,  Kesteven M.,  Wellington K.,   Roger R.,
  1987, \mn@doi [Aust. J. Phys.] {10.1071/ph870709}, 40, 709

\bibitem[\protect\citeauthoryear{Miville-Desch{\^{e}}nes, Ysard, Lavabre,
  Ponthieu, Mac{\'{I}}as-P{\'{e}}rez, Aumont  \&
  Bernard}{Miville-Desch{\^{e}}nes et~al.}{2008}]{MivilleDeschenes2008}
Miville-Desch{\^{e}}nes M.~A.,  Ysard N.,  Lavabre A.,  Ponthieu N.,
  Mac{\'{I}}as-P{\'{e}}rez J.~F.,  Aumont J.,   Bernard J.~P.,  2008, \mn@doi
  [A{\&}A] {10.1051/0004-6361:200809484}, 490, 1093

\bibitem[\protect\citeauthoryear{Momjian}{Momjian}{2017}]{Momjian2017}
Momjian E.,  2017, {Resolution — Science Website}, \url
  {https://science.nrao.edu/facilities/vla/docs/manuals/oss/performance/resolution}

\bibitem[\protect\citeauthoryear{Morlino \& Caprioli}{Morlino \&
  Caprioli}{2012}]{Morlino2012}
Morlino G.,  Caprioli D.,  2012, in AIP Conference Proceedings. pp 241--244,
  \mn@doi{10.1063/1.4772242}

\bibitem[\protect\citeauthoryear{Mulcahy et~al.,}{Mulcahy
  et~al.}{2014}]{Mulcahy2014}
Mulcahy D.~D.,  et~al., 2014, \mn@doi [A{\&}A] {10.1051/0004-6361/201424187},
  568

\bibitem[\protect\citeauthoryear{Murray, Quataert  \& Thompson}{Murray
  et~al.}{2005}]{Murray2005}
Murray N.,  Quataert E.,   Thompson T.~A.,  2005, \mn@doi [ApJ]
  {10.1086/426067}, 618, 569

\bibitem[\protect\citeauthoryear{Naab \& Ostriker}{Naab \&
  Ostriker}{2017}]{Naab2016}
Naab T.,  Ostriker J.~P.,  2017, \mn@doi [ARA{\&}A]
  {10.1146/annurev-astro-081913-040019}, 55, 59

\bibitem[\protect\citeauthoryear{Nava, Gabici, Marcowith, Morlino  \&
  Ptuskin}{Nava et~al.}{2016}]{Nava2016}
Nava L.,  Gabici S.,  Marcowith A.,  Morlino G.,   Ptuskin V.~S.,  2016,
  \mn@doi [MNRAS] {10.1093/mnras/stw1592}, 461, 3552

\bibitem[\protect\citeauthoryear{Ohno \& Shibata}{Ohno \&
  Shibata}{1993}]{Ohno1993}
Ohno H.,  Shibata S.,  1993, \mn@doi [MNRAS] {10.1093/mnras/262.4.953}, 262,
  953

\bibitem[\protect\citeauthoryear{Oppenheimer, Dav{\'{e}}, Kere{\v{s}}, Fardal,
  Katz, Kollmeier  \& Weinberg}{Oppenheimer et~al.}{2009}]{Oppenheimer2009}
Oppenheimer B.~D.,  Dav{\'{e}} R.,  Kere{\v{s}} D.,  Fardal M.,  Katz N.,
  Kollmeier J.~A.,   Weinberg D.~H.,  2009, \mn@doi [MNRAS]
  {10.1111/j.1365-2966.2010.16872.x}, 406, 2325

\bibitem[\protect\citeauthoryear{Pais, Pfrommer, Ehlert  \& Pakmor}{Pais
  et~al.}{2018}]{Pais2018a}
Pais M.,  Pfrommer C.,  Ehlert K.,   Pakmor R.,  2018, \mn@doi [MNRAS]
  {10.1093/mnras/sty1410}, 478, 5278

\bibitem[\protect\citeauthoryear{Pakmor, Pfrommer, Simpson  \& Springel}{Pakmor
  et~al.}{2016}]{Pakmor2016}
Pakmor R.,  Pfrommer C.,  Simpson C.~M.,   Springel V.,  2016, \mn@doi [ApJ]
  {10.3847/2041-8205/824/2/l30}, 824, L30

\bibitem[\protect\citeauthoryear{Persic \& Rephaeli}{Persic \&
  Rephaeli}{2010}]{Persic2010}
Persic M.,  Rephaeli Y.,  2010, \mn@doi [MNRAS]
  {10.1111/j.1365-2966.2009.16218.x}, 403, 1569

\bibitem[\protect\citeauthoryear{Pfrommer, En{\ss}lin, Springel, Jubelgas  \&
  Dolag}{Pfrommer et~al.}{2007}]{Pfrommer2007}
Pfrommer C.,  En{\ss}lin T.~A.,  Springel V.,  Jubelgas M.,   Dolag K.,  2007,
  \mn@doi [MNRAS] {10.1111/j.1365-2966.2007.11732.x}, 378, 385

\bibitem[\protect\citeauthoryear{Pfrommer, Pakmor, Schaal, Simpson  \&
  Springel}{Pfrommer et~al.}{2017}]{Pfrommer2017}
Pfrommer C.,  Pakmor R.,  Schaal K.,  Simpson C.~M.,   Springel V.,  2017,
  \mn@doi [MNRAS] {10.1093/mnras/stw2941}, 465, 4500

\bibitem[\protect\citeauthoryear{Pillepich et~al.,}{Pillepich
  et~al.}{2018}]{Pillepich2018}
Pillepich A.,  et~al., 2018, \mn@doi [MNRAS] {10.1093/mnras/stx2656}, 473, 4077

\bibitem[\protect\citeauthoryear{Pittard}{Pittard}{2013}]{Pittard2013}
Pittard J.,  2013, \mn@doi [MNRAS] {10.1093/mnras/stt1552}, 435, 3600

\bibitem[\protect\citeauthoryear{Polderman, Haverkorn  \& Jaffe}{Polderman
  et~al.}{2020}]{Polderman2020}
Polderman I.~M.,  Haverkorn M.,   Jaffe T.~R.,  2020, \mn@doi [A{\&}A]
  {10.1051/0004-6361/201937042}, 636, A2

\bibitem[\protect\citeauthoryear{Rand \& Kulkarni}{Rand \&
  Kulkarni}{1989}]{Rand1989}
Rand R.~J.,  Kulkarni S.~R.,  1989, \mn@doi [ApJ] {10.1086/167747}, 343, 760

\bibitem[\protect\citeauthoryear{Reich}{Reich}{2002}]{Reich2002}
Reich W.,  2002, in Becker W.,  Lesch H.,   Tr{\"{u}}mper J.,  eds, Neutron
  Stars, Pulsars, and Supernova Remnants. Max-Plank-Institut f{\"{u}}r
  extraterrestrische Physik, Garching bei M{\"{u}}nchen, \url
  {https://ui.adsabs.harvard.edu/abs/2002nsps.conf....1R}

\bibitem[\protect\citeauthoryear{Reissl, Wolf  \& Brauer}{Reissl
  et~al.}{2016}]{Reissl2016}
Reissl S.,  Wolf S.,   Brauer R.,  2016, \mn@doi [A{\&}A]
  {10.1051/0004-6361/201424930}, 593, 87

\bibitem[\protect\citeauthoryear{Reissl, Brauer, Klessen  \& Pellegrini}{Reissl
  et~al.}{2019}]{Reissl2019}
Reissl S.,  Brauer R.,  Klessen R.~S.,   Pellegrini E.~W.,  2019, \mn@doi [ApJ]
  {10.3847/1538-4357/ab3664}, 885, 15

\bibitem[\protect\citeauthoryear{Reynolds \& Gilmore}{Reynolds \&
  Gilmore}{1993}]{Reynolds1993}
Reynolds S.~P.,  Gilmore D.~M.,  1993, \mn@doi [AJ] {10.1086/116635}, 106, 272

\bibitem[\protect\citeauthoryear{Reynoso, Moffett, Goss, Dubner, Dickel,
  Reynolds  \& Giacani}{Reynoso et~al.}{1997}]{Reynoso1997}
Reynoso E.~M.,  Moffett D.~A.,  Goss W.~M.,  Dubner G.~M.,  Dickel J.~R.,
  Reynolds S.~P.,   Giacani E.~B.,  1997, \mn@doi [ApJ] {10.1086/304997}, 491,
  816

\bibitem[\protect\citeauthoryear{Reynoso, Hughes  \& Moffett}{Reynoso
  et~al.}{2013}]{Reynoso2013}
Reynoso E.~M.,  Hughes J.~P.,   Moffett D.~A.,  2013, \mn@doi [AJ]
  {10.1088/0004-6256/145/4/104}, 145, 104

\bibitem[\protect\citeauthoryear{Reynoso, Vel{\'{a}}zquez  \&
  Cichowolski}{Reynoso et~al.}{2018}]{Reynoso2018}
Reynoso E.~M.,  Vel{\'{a}}zquez P.~F.,   Cichowolski S.,  2018, \mn@doi [MNRAS]
  {10.1093/mnras/sty751}, 477, 2087

\bibitem[\protect\citeauthoryear{Rosdahl et~al.,}{Rosdahl
  et~al.}{2018}]{Rosdahl2018}
Rosdahl J.,  et~al., 2018, \mn@doi [MNRAS] {10.1093/mnras/sty1655}, 479, 994

\bibitem[\protect\citeauthoryear{Rosen \& Bregman}{Rosen \&
  Bregman}{1995}]{Rosen1995}
Rosen A.,  Bregman J.~N.,  1995, \mn@doi [ApJ] {10.1086/175303}, 440, 634

\bibitem[\protect\citeauthoryear{Ruszkowski, Yang  \& Zweibel}{Ruszkowski
  et~al.}{2017}]{Ruszkowski2017}
Ruszkowski M.,  Yang H.-Y.~K.,   Zweibel E.,  2017, \mn@doi [ApJ]
  {10.3847/1538-4357/834/2/208}, 834, 208

\bibitem[\protect\citeauthoryear{Salem, Bryan  \& Hummels}{Salem
  et~al.}{2014}]{Salem2014}
Salem M.,  Bryan G.~L.,   Hummels C.,  2014, \mn@doi [ApJ]
  {10.1088/2041-8205/797/2/L18}, 797, L18

\bibitem[\protect\citeauthoryear{Schekochihin, Cowley, Hammett, Maron  \&
  McWilliams}{Schekochihin et~al.}{2002}]{Schekochihin2002}
Schekochihin A.~A.,  Cowley S.~C.,  Hammett G.~W.,  Maron J.~L.,   McWilliams
  J.~C.,  2002, \mn@doi [New J. Phys.] {10.1088/1367-2630/4/1/384}, 4, 84

\bibitem[\protect\citeauthoryear{Schroer, Pezzi, Caprioli, Haggerty  \&
  Blasi}{Schroer et~al.}{2021}]{Schroer2020}
Schroer B.,  Pezzi O.,  Caprioli D.,  Haggerty C.,   Blasi P.,  2021, \mn@doi
  [ApJ] {10.3847/2041-8213/ac02cd}, 914, L13

\bibitem[\protect\citeauthoryear{Sedov}{Sedov}{1959}]{Sedov1959}
Sedov L.~I.,  1959, {Similarity and Dimensional Methods in Mechanics}.
Elsevier, \mn@doi{10.1016/C2013-0-08173-X}

\bibitem[\protect\citeauthoryear{Semenov, Kravtsov  \& Gnedin}{Semenov
  et~al.}{2018}]{Semenov2018}
Semenov V.~A.,  Kravtsov A.~V.,   Gnedin N.~Y.,  2018, \mn@doi [ApJ]
  {10.3847/1538-4357/aac6eb}, 861, 4

\bibitem[\protect\citeauthoryear{Semenov, Kravtsov  \& Caprioli}{Semenov
  et~al.}{2021}]{Semenov2021}
Semenov V.~A.,  Kravtsov A.~V.,   Caprioli D.,  2021, \mn@doi [ApJ]
  {10.3847/1538-4357/abe2a6}, 910, 126

\bibitem[\protect\citeauthoryear{Sijacki, Pfrommer, Springel  \&
  En{\ss}lin}{Sijacki et~al.}{2008}]{Sijacki2008}
Sijacki D.,  Pfrommer C.,  Springel V.,   En{\ss}lin T.~A.,  2008, \mn@doi
  [MNRAS] {10.1111/j.1365-2966.2008.13310.x}, 387, 1403

\bibitem[\protect\citeauthoryear{Smith, Sijacki  \& Shen}{Smith
  et~al.}{2018}]{Smith2018}
Smith M.~C.,  Sijacki D.,   Shen S.,  2018, \mn@doi [MNRAS]
  {10.1093/mnras/sty994}, 478, 302

\bibitem[\protect\citeauthoryear{Smith, Sijacki  \& Shen}{Smith
  et~al.}{2019}]{Smith2019}
Smith M.~C.,  Sijacki D.,   Shen S.,  2019, \mn@doi [MNRAS]
  {10.1093/mnras/stz599}, 485, 3317

\bibitem[\protect\citeauthoryear{Somerville \& Primack}{Somerville \&
  Primack}{1999}]{Somerville1999}
Somerville R.~S.,  Primack J.~R.,  1999, \mn@doi [MNRAS]
  {10.1046/j.1365-8711.1999.03032.x}, 310, 1087

\bibitem[\protect\citeauthoryear{Springel \& Hernquist}{Springel \&
  Hernquist}{2003}]{Springel2003}
Springel V.,  Hernquist L.,  2003, \mn@doi [MNRAS]
  {10.1046/j.1365-8711.2003.06207.x}, 339, 312

\bibitem[\protect\citeauthoryear{Stinson, Seth, Katz, Wadsley, Governato  \&
  Quinn}{Stinson et~al.}{2006}]{Stinson2006}
Stinson G.,  Seth A.,  Katz N.,  Wadsley J.,  Governato F.,   Quinn T.,  2006,
  \mn@doi [MNRAS] {10.1111/j.1365-2966.2006.11097.x}, 373, 1074

\bibitem[\protect\citeauthoryear{Sun, Reich, Waelkens  \& En{\ss}lin}{Sun
  et~al.}{2008}]{Sun2008}
Sun X.~H.,  Reich W.,  Waelkens A.,   En{\ss}lin T.~A.,  2008, \mn@doi [A{\&}A]
  {10.1051/0004-6361:20078671}, 477, 573

\bibitem[\protect\citeauthoryear{Tasker}{Tasker}{2011}]{Tasker2011}
Tasker E.~J.,  2011, \mn@doi [ApJ] {10.1088/0004-637X/730/1/11}, 730

\bibitem[\protect\citeauthoryear{Teyssier}{Teyssier}{2002}]{Teyssier2002}
Teyssier R.,  2002, \mn@doi [A{\&}A] {10.1051/0004-6361:20011817}, 385, 337

\bibitem[\protect\citeauthoryear{Teyssier, Pontzen, Dubois  \& Read}{Teyssier
  et~al.}{2013}]{Teyssier2013}
Teyssier R.,  Pontzen A.,  Dubois Y.,   Read J.~I.,  2013, \mn@doi [MNRAS]
  {10.1093/mnras/sts563}, 429, 3068

\bibitem[\protect\citeauthoryear{Thornton, Gaudlitz, Janka  \&
  Steinmetz}{Thornton et~al.}{1998}]{Thornton1998}
Thornton K.,  Gaudlitz M.,  Janka H.,   Steinmetz M.,  1998, \mn@doi [ApJ]
  {10.1086/305704}, 500, 95

\bibitem[\protect\citeauthoryear{Thoudam \& H{\"{o}}randel}{Thoudam \&
  H{\"{o}}randel}{2012}]{Thoudam2012}
Thoudam S.,  H{\"{o}}randel J.~R.,  2012, \mn@doi [MNRAS]
  {10.1111/j.1365-2966.2011.20385.x}, 421, 1209

\bibitem[\protect\citeauthoryear{Turk, Smith, Oishi, Skory, Skillman, Abel  \&
  Norman}{Turk et~al.}{2011}]{Turk2011}
Turk M.~J.,  Smith B.~D.,  Oishi J.~S.,  Skory S.,  Skillman S.~W.,  Abel T.,
  Norman M.~L.,  2011, \mn@doi [ApJS] {10.1088/0067-0049/192/1/9}, 192, 9

\bibitem[\protect\citeauthoryear{Uhlig, Pfrommer, Sharma, Nath, En{\ss}lin  \&
  Springel}{Uhlig et~al.}{2012}]{Uhlig2012}
Uhlig M.,  Pfrommer C.,  Sharma M.,  Nath B.~B.,  En{\ss}lin T.~A.,   Springel
  V.,  2012, \mn@doi [MNRAS] {10.1111/j.1365-2966.2012.21045.x}, 423, 2374

\bibitem[\protect\citeauthoryear{Veilleux, Cecil  \& Bland-Hawthorn}{Veilleux
  et~al.}{2005}]{Veilleux2005}
Veilleux S.,  Cecil G.,   Bland-Hawthorn J.,  2005, \mn@doi [ARA{\&}A]
  {10.1146/annurev.astro.43.072103.150610}, 43, 769

\bibitem[\protect\citeauthoryear{Vink, Aharonian, Hofmann  \& Rieger}{Vink
  et~al.}{2008}]{Vink2009}
Vink J.,  Aharonian F.~A.,  Hofmann W.,   Rieger F.,  2008, in AIP Conference
  Proceedings. AIP, pp 169--180, \mn@doi{10.1063/1.3076632}, \url
  {http://aip.scitation.org/doi/abs/10.1063/1.3076632}

\bibitem[\protect\citeauthoryear{Vishniac}{Vishniac}{1983}]{Vishniac1983}
Vishniac E.~T.,  1983, \mn@doi [ApJ] {10.1086/161433}, 274, 152

\bibitem[\protect\citeauthoryear{Vogelsberger, Genel, Sijacki, Torrey, Springel
   \& Hernquist}{Vogelsberger et~al.}{2013}]{Vogelsberger2013}
Vogelsberger M.,  Genel S.,  Sijacki D.,  Torrey P.,  Springel V.,   Hernquist
  L.,  2013, \mn@doi [MNRAS] {10.1093/mnras/stt1789}, 436, 3031

\bibitem[\protect\citeauthoryear{Vogelsberger, Marinacci, Torrey  \&
  Puchwein}{Vogelsberger et~al.}{2020}]{Vogelsberger2020}
Vogelsberger M.,  Marinacci F.,  Torrey P.,   Puchwein E.,  2020, \mn@doi [Nat.
  Rev. Phys.] {10.1038/s42254-019-0127-2}, 2, 42

\bibitem[\protect\citeauthoryear{Wadepuhl \& Springel}{Wadepuhl \&
  Springel}{2011}]{Wadepuhl2011}
Wadepuhl M.,  Springel V.,  2011, \mn@doi [MNRAS]
  {10.1111/j.1365-2966.2010.17576.x}, 410, 1975

\bibitem[\protect\citeauthoryear{Walch \& Naab}{Walch \&
  Naab}{2015}]{Walch2015}
Walch S.,  Naab T.,  2015, \mn@doi [MNRAS] {10.1093/mnras/stv1155}, 451, 2757

\bibitem[\protect\citeauthoryear{Webber}{Webber}{1998}]{Webber1998}
Webber W.~R.,  1998, \mn@doi [ApJ] {10.1086/306222}, 506, 329

\bibitem[\protect\citeauthoryear{West, Safi-Harb, Jaffe, Kothes, Landecker  \&
  Foster}{West et~al.}{2016}]{West2015}
West J.~L.,  Safi-Harb S.,  Jaffe T.,  Kothes R.,  Landecker T.~L.,   Foster
  T.,  2016, \mn@doi [A{\&}A] {10.1051/0004-6361/201527001}, 587, A148

\bibitem[\protect\citeauthoryear{West, Jaffe, Ferrand, Safi-Harb  \&
  Gaensler}{West et~al.}{2017}]{West2017}
West J.~L.,  Jaffe T.,  Ferrand G.,  Safi-Harb S.,   Gaensler B.~M.,  2017,
  \mn@doi [ApJ] {10.3847/2041-8213/aa94c4}, 849, L22

\bibitem[\protect\citeauthoryear{Zweibel}{Zweibel}{2017}]{Zweibel2017}
Zweibel E.~G.,  2017, \mn@doi [Phys. Plasmas] {10.1063/1.4984017}, 24

\makeatother
\end{thebibliography}




\appendix
\section{Case example of momentum modelling}
\label{ap:example_momentum_model}
\begin{figure}
  \centering \includegraphics[width=\columnwidth]{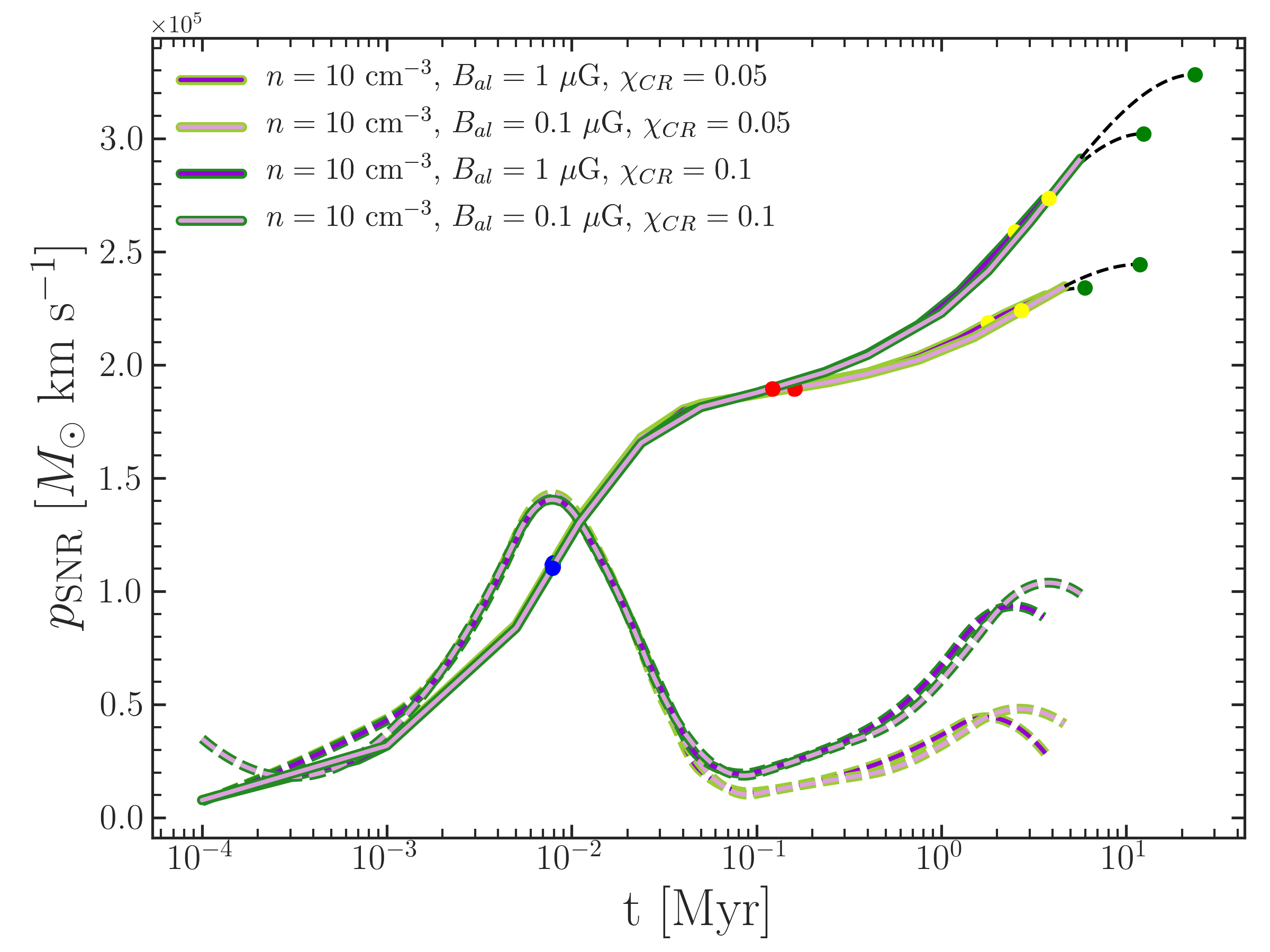}
  \caption{Evolution of the total outward radial momentum of the SNR (solid lines) with time for the CRMHD simulations with $n=10$ cm$^{-3}$, $\vert \vec{B} \vert=0.1$ and 1 $\mu$G, and $\chi_{\text{CR}}=0.05$ and 0.1. Colour scheme is as for Fig. \ref{fig:momentum_vs_t}. Coloured dashed lines are the temporal derivative of $p_{\text{SNR}}$ and the black dashed lines indicate the extrapolation until the point of convergence in our model. Circles plotted on top of the simulation data correspond to important moments in the momentum evolution: maximum growth rate during Sedov-Taylor (blue), surpassing of HD momentum by each CRMHD run (red), final inflection point towards convergence (yellow), and final maximum momentum (green). Even when a momentum plateau is not reached during the time over which we run the simulation, extending their evolution after the final inflection point can be used to predict a smooth transition towards the final value. There is only a slight dependence of the final momentum and convergence time on the strength of the aligned magnetic field.}
  \label{fig:model}
\end{figure}

We show in Fig. \ref{fig:model} the full evolution of radial outward momentum for the CRMHD runs with 0.1 and 1 $\mu$G, and $\chi_{\rm CR} = 0.05$ and 0.1 to at least 4 Myr and up to 6 Myr and normalised for clarity. This is given by the solid coloured lines. These showcase magnetic field strength increasing with darker shades of purple and SNe energy fraction in CRs as contours with darker shades of green. The figure also shows the temporal derivative of this data as the dashed lines. The first significant point we search for is that when the maximum growth rate of the Sedov-Taylor phase is reached. We have shown that the addition of CRs does not alter the evolution with respect to the HD runs during the classical Sedov-Taylor phase despite the transference of SN thermal energy to CR energy. Hence, the growth during the Sedov-Taylor phase should not depend neither on the strength of the magnetic field nor the amount of CR energy for a given ambient density. The maximum in the momentum time derivative is represented by blue points, which lie in top of each other, as predicted. Furthermore, the point when the momentum gain of the CRMHD runs exceeds that of the HD runs happens earlier for higher $\chi_{\rm CR}$ (see Section \ref{subsec:momentum_hd_mhd}). This is shown by the red points in Fig. \ref{fig:model}. After this, our algorithm looks for the last point of inflection, which is depicted here by the yellow points. The simulation data after these points does not always show clear convergence of momentum. For these cases, we make use of the spline interpolation to extend the evolution of $p_{\rm SNR}$ up until its maximum value. This is the time of  plateauing, shown by the green points (the information regarding which simulations have reached this point is provided in Table \ref{tab:sim_list}). We evolved our 0.1$\mu$G runs for 2 additional Myr to find the inflection point, which took place at a later time than for the 1 $\mu$G runs. Consequently, runs with a lower magnetic field appear to converge at a later time and with a slightly higher momentum. We attribute this difference to the effects of the magnetised medium on the morphology of the shell: when a higher magnetic field is present, the widening of the shell is enhanced, and the thinner shell at late times is less capable of confining CRs for their additional momentum deposition. Nevertheless, we find this difference to be below $\sim 10$\%. In order to keep our CR model as simple as possible, we ignore this minor dependence on magnetic field.
\section{Separating regions of the SNR}
\label{ap:region_algorithm}
SNRs have been studied as a system in which the interplay of three distinct regions was fundamental in its understanding: the inner bubble, the thin shell and the surrounding medium. Traditionally, a separation based in temperature and radial velocity has been used \citep[e.g][]{Kim2015}, supported by the details inferred by analytical results. However, magnetic fields and CRs have different thermodynamical effects depending on many of the initial parameters that we consider (e.g. ambient density, CR energy fraction, magnetic field configuration), so setting hard temperature and radial thresholds would be a way of biasing our results. Alternatively, a geometrical separation procedure would be suitable in the case of a near-perfect spherical symmetry, but, as we have detailed in our results, that is far from being the norm (e.g. shell-widening in MHD and CRMHD runs with aligned magnetic field). Considering these subtleties, we have developed two simple algorithms that separate the different regions of the SNR. The Method I is used for simulations that have near-spherical symmetry (i.e. HD and runs with tangled initial magnetic field):
\begin{enumerate}
    \item Using the average radial profiles in the $x$ and $z$ directions from Section~\ref{subsec:profiles_hd_mhd}, compute the second derivative of the density profile.
    \item Search for the changes in sign of the second derivative (i.e. inflection points) in order to find the outer $r_{\rm out}$ and inner $r_{\rm in}$ limits of the shell. Widen those limits by 10\% to allow for instabilities and magnetically-induced perturbations to be capture as shell.
    \item Select the bubble as: a) all gas with density below the average measured at $r_{\rm in}$ and with radial distance larger than $r_{\rm in}$ and smaller than $r_{\rm out}$, or b) all gas below $r_{\rm in}$.
    \item The ambient gas is selected in the opposite way to the bubble: a) all gas with density above the average measured at $r_{\rm in}$ and with radial distance larger than $r_{\rm in}$ and smaller than $r_{\rm out}$, or b) all gas above $r_{\rm out}$.
    \item Shell is selected as all gas between $r_{\rm in}$ and $r_{\rm out}$ with density higher than the average the average measured at $r_{\rm in}$, with radial velocities $> 1$ km/s.
\end{enumerate}
\begin{figure}
  \centering \includegraphics[width=\columnwidth]{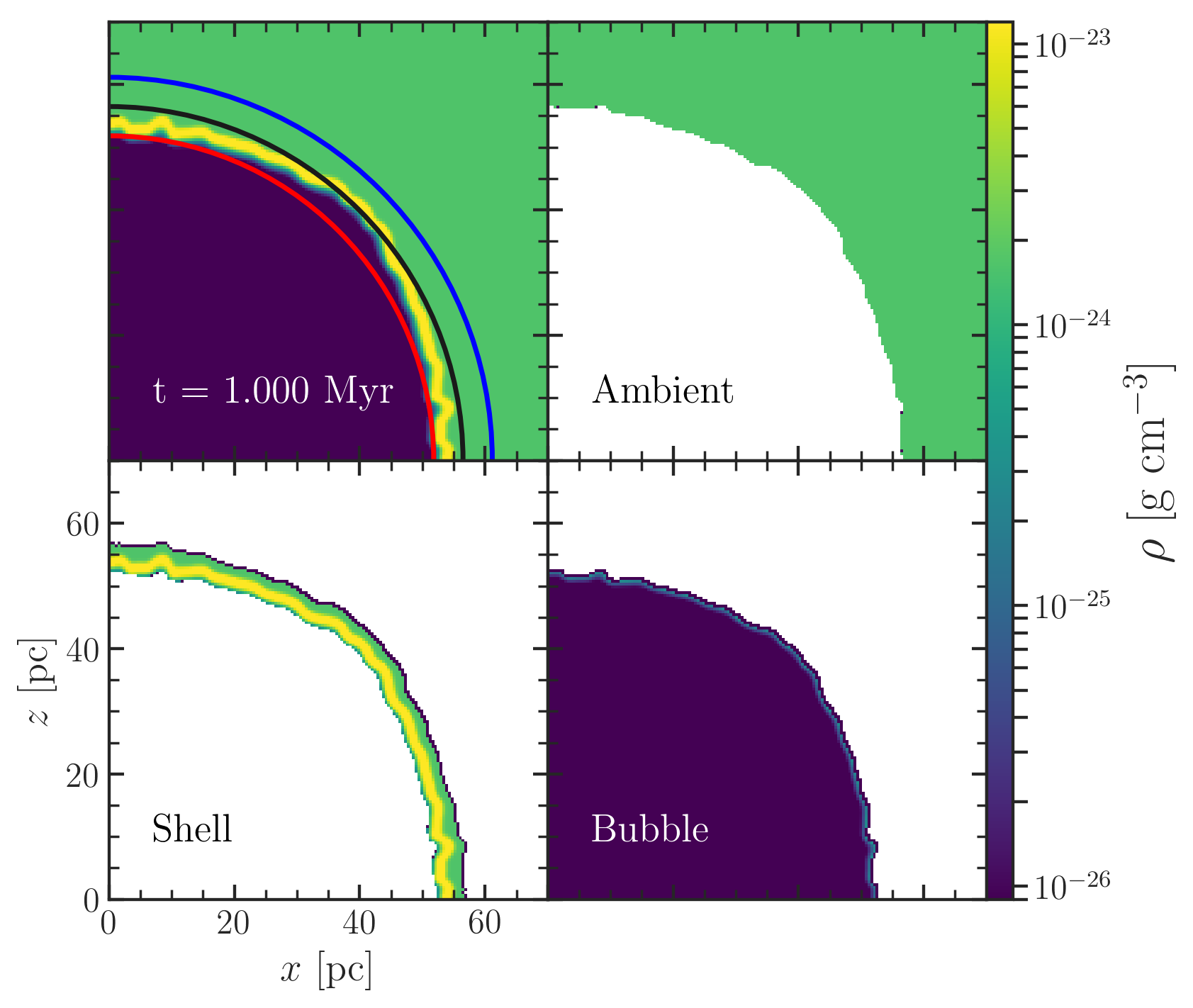}
  \caption{Slices through the $XZ$ plane of the HD run regions (i.e. ambient medium, shell and bubble) with $n=1$ cm$^{-3}$ at $1$~Myr, showing the top-right quadrant of the SNR. In the top left panel, $r_{\rm in}$ (red circle) and $r_{\rm out}$ (blue circle) from the Method I of region selection. The rest of the panels, in counterclockwise order, are: ambient medium, bubble and shell.}
  \label{fig:region_example}
\end{figure}
Fig.~\ref{fig:region_example} shows the results of this algorithm for the HD run with $n=1$ cm$^{-3}$. In the top left panel, $r_{\rm in}$ (red circle) and $r_{\rm out}$ (blue circle) are plotted in top of a thin slice of the top right quadrant of the SNR. The solid black line indicates the average radial distance of the shell, showing how instabilities make the bounding of the shell from below and above a difficult task. The rest of the panels show the selected ambient medium, shell and bubble.
The Method II is used for runs presenting axisymmetry (i.e. with an initially uniform magnetic field):
\begin{enumerate}
    \item As for the previous method, using the average radial density profiles in the $x$ and $z$ directions, search for the minimum gas density $\rho_{\rm min}$ which is inside the SNR (i.e. radial distance smaller than the peak in density of the shell $r_{\rm peak}$).
    \item By computing the second derivative of the density profile, define $r_{\rm out}$ as the closest inflection point to $r_{\rm peak}$ with $r>r_{\rm peak}$.
    \item Select the bubble as everything below $r_{\rm out}$ with density below twice $\rho_{\rm min}$. We use this factor of 2 after finding that the majority of MHD and CRMHD runs with a uniform magnetic field present that value of density at the inner inflection point (i.e. limit between bubble and shell). We have explored variations of these factor, finding our results transparent to this parameter. Only small variations in bubble volume are found ($\sim 5$\%).
    \item The shell canonical shell is selected as everything below $r_{\rm out}$ with density above or equal than twice $\rho_{\rm min}$. Additionally, in order to select the outflows found in Section~\ref{subsec:results_hd_mhd}, we add to the shell the regions with density above or equal than twice $\rho_{\rm min}$ travelling in the radial direction with velocities $>1$ km/s.
\end{enumerate}

\bsp	
\label{lastpage}
\end{document}